\documentclass[%
 aip,
 amsmath,amssymb,
 reprint,%
]{revtex4-1}
\usepackage{graphicx}
\usepackage{dcolumn}
\usepackage{bm}
\usepackage[utf8]{inputenc}
\usepackage[T1]{fontenc}
\usepackage{mathptmx}
\usepackage{etoolbox}
\makeatletter
\def\@email#1#2{%
 \endgroup
 \patchcmd{\titleblock@produce}
  {\frontmatter@RRAPformat}
  {\frontmatter@RRAPformat{\produce@RRAP{*#1\href{mailto:#2}{#2}}}\frontmatter@RRAPformat}
  {}{}
}%
\makeatother

\begin{document}

\preprint{AIP/123-QED}

\title{Causal self-consistency of the Blandford--McKee self-similar solution}

\author{Gilad Sadeh}
\altaffiliation[]{gilad.sadeh@aei.mpg.de}
\affiliation{Max Planck Institute for Gravitational Physics (Albert Einstein Institute), Am Mühlenberg 1,
Potsdam-Golm, 14476, Germany}
\affiliation{Dept. of Particle Phys. \& Astrophys., Weizmann Institute of Science, Rehovot 76100, Israel}

\author{Tamar Faran}
\affiliation{Dept. of Particle Phys. \& Astrophys., Weizmann Institute of Science, Rehovot 76100, Israel}

\author{Doron Kushnir}
\affiliation{Dept. of Particle Phys. \& Astrophys., Weizmann Institute of Science, Rehovot 76100, Israel}

\author{Eli Waxman}
\affiliation{Dept. of Particle Phys. \& Astrophys., Weizmann Institute of Science, Rehovot 76100, Israel}

\date{\today}

\begin{abstract}
The Blandford--McKee (BM) solution describes the ultra-relativistic self-similar flow behind a strong spherical blast wave propagating into an external density profile $\propto r^{-k}$, where $k<4$ and $r$ is the distance from the center, but it applies only to the hot shell adjacent to the shock, sufficiently deep behind it, the fluid leaves the BM regime. Since the similarity profiles are determined solely by the shock conditions, with no conditions imposed at the inner end, the validity of the solution near the shock depends on whether the flow beyond the BM regime can imprint on this hot shell. For shallow density profiles, $k<k_g\simeq 2.062$, the characteristic structure of the BM solution already prevents forward-going acoustic information from the non-BM interior from reaching the shock during the ultra-relativistic stage. For steeper profiles, $k_g<k<4$, the hot equation of state fails while the flow is still relativistic and shock-connected. There, we derive a new self-similar solution for the cooling relativistic flow, which has its own similarity scale, equations, and characteristic structure. It overlaps with the hot BM solution toward the shock, and reaches a sonic point beyond this overlap. This critical point separates the shock-connected BM-plus-cooling composite from the deeper downstream region, so acoustic signals generated beyond it cannot propagate toward the shock. The mechanism is explicit at $k=7/2$, for which the cooling solution is obtained analytically. This causal structure is therefore what enables the BM solution, which remains self-consistent near the shock even though it does not globally describe the full downstream flow. 
\end{abstract}

\maketitle

\section{Introduction}
Self-similar solutions play a central role in shock dynamics, offering powerful mathematical simplification by reducing partial differential equations with spatial and temporal derivatives to ordinary differential equations through a single similarity variable. They characterize the asymptotic behavior of physical systems once the influence of initial scales becomes negligible and, in some cases, can even be expressed analytically. Beyond their mathematical elegance, self-similar solutions provide clear physical insight into complex systems. One of the classic manifestations of self-similarity in hydrodynamics is the evolution of a blast wave following a powerful explosion. Suppose the surrounding external medium has a power-law density profile, $\rho_\text{ext}\propto r^{-\omega}$, with $\omega$ a constant (following the conventions of the respective literature, we denote the power-law index by \(\omega\) in the non-relativistic problem and by \(k\) in the relativistic one). In the non-relativistic case, the injection of energy at the origin launches a strong shock whose radius grows in time as $R(t)\propto t^\alpha$ ($t$ is the time since explosion), where the exponent $\alpha$ depends on $\omega$. When dimensional analysis or global conservation laws are enough to determine $\alpha$, this family of solutions is classified as self-similarity of the first type. This is the famous Sedov-von Neumann-Taylor solution.\citep{sedov_propagation_1946,von_neumann_point_1963,taylor_formation_1950} 
The situation changes if the ambient density falls off too steeply. It was demonstrated that for $\omega > 3$ the dimensional estimate of $\alpha$ breaks down:\citep{waxman_secondtype_1993} the shock accelerates, outrunning the flow behind it and losing causal connection. In such cases, the correct scaling is fixed not by dimensional reasoning but by demanding that the solution pass smoothly through a singular point of the governing equations, corresponding to the sonic point. These are known as second-type solutions and occur for $\omega >\omega_g\simeq 3.26$ (for the intermediate regime, $3<\omega<\omega_g$, a different solution applies \citep{gruzinov_self-similarity_2003,kushnir_closing_2010}). 

The ultra-relativistic analog of the Sedov-von Neumann-Taylor solution was derived by Blandford and McKee (BM).\citep{blandford_fluid_1976} They derived a first-type self-similar solution describing a spherical ultra-relativistic blast wave expanding into an external medium with $\rho_\text{ext}\propto r^{-k}$ (a corresponding second-type relativistic analog also exists\citep{best_second-type_2000,wang_stability_2003,sari_first_2006}). In this framework, the shock Lorentz factor evolves as $\Gamma^2\propto t^{-m}$ (where the exponent $m$ depends on $k$), and the post-shock region follows a family of similarity profiles. The BM derivation rests on two assumptions: (i) the shocked fluid remains ultra-relativistic throughout the flow, (ii)
an ultra-relativistic equation of state (EoS), $p=\varepsilon/3$ (radiation‑dominated, $p$ and $\varepsilon$ are the fluid pressure and total proper energy density, respectively). These two assumptions break down in the interior of the flow. For moderate density profiles ($k<2$), the inner layers become trans-relativistic before significant cooling begins,\citep{faran_non-relativistic_2021} while for steep density profiles ($k>2$), inner layers of the shocked fluid can remain kinematically relativistic while becoming thermodynamically trans-relativistic ($p/\rho c^{2} \lesssim 1$, where $\rho$ is the shocked fluid proper mass density), 
rest-mass inertia becomes dynamically important, the sound speed drops below its hot relativistic value, and the hot EoS, $p=\varepsilon/3$, is no longer adequate. The hot BM branch is therefore not globally valid: a different asymptotic description is required for the deeper downstream flow.\citep{faran_non-relativistic_2021} This exposes a gap in the derivation itself. The BM construction is global in form but local in validity: the similarity form is imposed
on the entire flow behind the shock, and the profiles follow from the shock conditions alone, with no condition imposed at the inner end, yet the underlying assumptions fail at a finite depth. The derivation thus implicitly relies on the deeper flow being either self-similar as well or irrelevant to the shell, and neither is guaranteed a priori. If acoustic signals generated in the deeper region could propagate forward to the shock, the shock and the adjacent hot shell would not be dynamically closed, and it would not be clear why the near-shock part of the solution should be correct.

In this paper, we show that the flow itself closes this gap. We first analyze the characteristic structure of the hot BM solution and identify a critical density index $k_g=\frac{7\sqrt{3}}{2}-4\simeq2.062$. For $k<k_g$, forward-going characteristics launched from the non-self-similar interior do not reach the hot BM shell during the ultra-relativistic stage. In this regime, the hot BM characteristic structure alone is sufficient to causally protect the BM solution. For $k>k_g$, the thermodynamic transition out of the hot EoS occurs while the flow is still ultra-relativistic and causally connected to the shock. The hot BM characteristic argument is then insufficient, and the cooling layer must be resolved explicitly. We construct a cooling self-similar solution matched asymptotically to the hot BM shell, following the composite-solution approach of Pan \& Sari.\citep{pan_composite_2009} This solution has its own similarity scale, equations, and characteristic structure. The BM-connected cooling solution remains shock-connected near the hot overlap, but it reaches a sonic point before the deeper interior is encountered. In the reduced similarity equations, this occurs either at a finite-sound-speed sonic point or at a zero-sound-speed sonic point. The corresponding sonic point in the physical flow marks the inner boundary of the BM-connected composite solution. Beyond this point, forward acoustic signals cannot propagate back to the shock in similarity space. 

This identifies the physical significance of the cooling solution. Its importance is not merely to continue the BM profiles into the colder flow. Rather, it turns out to establish the
causal structure of the downstream region and, in particular, the acoustic barrier that shields the hot relativistic shell from the deeper interior (this property is not automatic and is geometry dependent, App.~\ref{app:geometry}). In this sense, the causal structure is not a consistency check on the BM solution but the property that enables it. A useful analytic benchmark is provided by the case $k=7/2$. In that case, the full cooling solution is obtained analytically, making the causal shielding mechanism completely explicit.

Ultra-relativistic shock waves are central to many high-energy astrophysical transients and power some of the most luminous explosions in the Universe. Owing to relativistic beaming and the limited lateral causal contact of ultra-relativistic outflows, a sufficiently narrow jet evolves, to leading order, as a portion of a spherical blast wave. The BM solution therefore provides the standard hydrodynamic framework for modeling gamma-ray burst afterglows.\citep{meszaros_optical_1997,waxman_gamma-ray--burst_1997,sari_spectra_1998,granot_shape_2002,sadeh_hydrodynamics_2026} Related relativistic shock models have also been applied to less relativistic transients, including the afterglow of the binary-neutron-star merger, GW170817,\citep{granot_off-axis_2018,ghirlanda_compact_2019,ryan_gamma-ray_2020,beniamini_afterglow_2020,ryan_modeling_2024} as well as to tidal disruption events \citep{beniamini_swift_2023,cendes_continued_2025} and luminous fast blue optical transients,\citep{nayana_most_2025} where the radio emission is commonly attributed to shock-powered outflows. These diverse applications highlight the central role of relativistic shocks in shaping the observed signatures of cosmic explosions.

The paper is organized as follows. In \S~\ref{sec:hot_bm} we review
the hot BM solution, identify where its asymptotic assumptions fail, and
derive the hot-BM characteristic criterion. In \S~\ref{sec:cooling} we construct the cooling self-similar solution and
analyze its characteristic structure, sonic points, and its implications. Finally, we summarize the results in \S~\ref{sec:summary}.

\section{The hot Blandford--McKee shell}
\label{sec:hot_bm}
The main symbols used throughout the paper are collected in App.~\ref{app:glossary}.
\subsection{The self-similar solution}
We begin by outlining the original BM self-similar solution for an adiabatic impulsive blast wave. In units of $c=1$, the relativistic conservative, spherical, hydrodynamic equations for a perfect fluid are
\begin{equation}
\begin{aligned}
\partial_t\!\left[\gamma^2(\varepsilon+\beta^2p)\right]
+\frac{1}{r^2}\partial_r\!\left[
r^2\gamma^2\beta(\varepsilon+p)\right]
&=0,\\
\partial_t\!\left[\gamma^2\beta(\varepsilon+p)\right]
+\frac{1}{r^2}\partial_r\!\left[
r^2\left(\gamma^2\beta^2(\varepsilon+p)+p\right)\right]-\frac{2p}{r}
&=0,\\
\partial_t(\gamma\rho)
+\frac{1}{r^2}\partial_r(r^2\gamma\beta\rho)
&=0,
\end{aligned}
\label{eq:relativistic_euler}
\end{equation}
where $\gamma$ and $\beta$ are the Lorentz factor and velocity of the fluid, respectively. The BM solution assumes an ultra-relativistic EoS, $p=\varepsilon/3$, and that the blast wave is ultra-relativistic, such
that the Lorentz factor of the shock front, \(\Gamma\), and that of the
shocked fluid, \(\gamma\), are both \(\gg1\). The solution is then
derived to lowest order in \(\Gamma^{-2}\) and \(\gamma^{-2}\). The external medium is assumed to be a scale-free power-law, $\rho_\text{ext}\propto r^{-k}$. Radiative losses are neglected, so the total energy contained within the shock remains constant. The shock Lorentz factor is assumed to evolve as
\begin{equation}
    \Gamma^2\propto t^{-m},\quad m>-1.
\end{equation}
To leading order in $\Gamma^{-2}$ the shock radius is
\begin{equation}
    R=t\left[1-\frac{1}{2(m+1)\Gamma^2}\right].
\end{equation}
The BM similarity variable is
\begin{equation}
\label{eq:chi_def}
\chi=1+2(m+1)\left(1-\frac{r}{R}\right)\Gamma^2
\simeq\left[1+2(m+1)\Gamma^2\right]\left(1-\frac{r}{t}\right).
\end{equation}
Behind the shock front, the pressure, Lorentz factor, and density in the shocked fluid, respectively, can be written in terms of $\chi$ as,
   \begin{equation}
   \label{eq:BM}
       \begin{aligned}
           p&=\frac{2}{3}\rho_\text{ext}(R)\Gamma^2f(\chi),\\
        \gamma^2&=\frac{1}{2}\Gamma^2g(\chi),\\
        \gamma\rho&=2\rho_\text{ext}(R)\Gamma^2h(\chi).
       \end{aligned}
   \end{equation}
where $\chi\geq1$ and the external medium is assumed to be cold. The boundary conditions for a strong ultra-relativistic shock are satisfied by
\begin{equation}
    f(1)=g(1)=h(1)=1,
\end{equation}
ensuring the shock jump conditions. For an impulsive adiabatic blast wave, energy conservation gives $m=3-k>-1$. Then, solving Eqs.~(\ref{eq:relativistic_euler}) leads to 
\begin{equation}
    \begin{aligned}
    \label{eq:BM_profiles}
       &f(\chi)=\chi^{\frac{4k-17}{12-3k}}\equiv\chi^{\alpha_f},\\
    &g(\chi)=\chi^{-1},\\
    &h(\chi)=\chi^{\frac{2k-7}{4-k}}\equiv\chi^{\alpha_h}. 
    \end{aligned}
\end{equation}

\subsection{Breakdown of the hot BM asymptotic regime}
\label{sec:break}
The BM solution is controlled by two local assumptions. Identifying
where each of them fails determines where the non-BM interior begins.
\begin{itemize}
\item First, the
post-shock flow must remain ultra-relativistic, so that the expansion in $\gamma^{-2}$ remains valid. Deeper in the flow, the fluid decelerates and approaches the non-relativistic interior.\citep{faran_non-relativistic_2021} Since $\gamma^2=\frac{1}{2\chi}\Gamma^2$, we define the kinematic trans-relativistic
cutoff at $\gamma_\text{tr}=3$ (order-unity convention), leading to
\begin{equation}
\label{eq:chi_slow}
\chi_\text{slow}=\frac{\Gamma^2}{18}.
    \end{equation} 
\item Second, the shocked fluid must remain thermodynamically relativistic, so
that $p/\rho\gg1$. Using the BM profiles, one finds \begin{equation}
\frac{p}{\rho}
=
\frac{f(\chi)\Gamma\sqrt{g(\chi)}}{3\sqrt{2}\,h(\chi)}
=
\frac{\Gamma}{3\sqrt{2}}\,
\chi^{(k+4)/[6(k-4)]}.
\end{equation}
We define the thermodynamic trans-relativistic cutoff by $p/\rho=1$, namely
\begin{equation}
\label{eq:chi_cold}
\chi_{\rm cold}
=
\left(\frac{\Gamma^2}{18}\right)^{\frac{3(4-k)}{k+4}}.
\end{equation}
\end{itemize} 
$\chi_{\rm cold}<\chi_{\rm slow}$ for $k>2$, thus, moving inward from the shock, the hot EoS fails before the flow becomes kinematically trans-relativistic. For $k<2$, the ordering is reversed: the flow becomes kinematically trans-relativistic before significant thermodynamic cooling.\citep{faran_non-relativistic_2021} It is useful to denote by $\chi_b$ the location where the hot BM asymptotic description first breaks down:
\begin{equation}
    \label{eq:chi_c}
        \chi_b=\begin{cases}
           \frac{\Gamma^2}{18} ,\quad &\text{for } k\leq2,\\
           \left(\frac{\Gamma^2}{18}\right)^{\frac{3(4-k)}{k+4}} ,\quad &\text{for } k>2.
        \end{cases}
    \end{equation}
    The value $\gamma_\text{tr}=3$ leading to Eq. \eqref{eq:chi_slow} can be replaced by any order-unity $\gamma_\text{tr}>1$ provided the thermodynamic cutoff is rescaled accordingly, $p/\rho = \gamma_\text{tr}/3$. This pairing keeps $\chi_b$ continuous at $k = 2$. The asymptotic results below depend only on the scalings $\chi_{\rm slow}\propto\Gamma^2$ and $\chi_{\rm cold}\propto\left(\Gamma^2\right)^{\frac{3(4-k)}{k+4}}$.
    
Since $\Gamma^2\propto t^{k-3}$ for the impulsive BM solution, we have
\begin{equation}
\frac{d\ln\chi_b}{d\ln t} =
\begin{cases}
k-3, & \text{for }k\leq 2,\\[4pt]
(k-3)\dfrac{3(4-k)}{k+4}, & \text{for }k>2.
\end{cases}
\end{equation}

\subsection{BM characteristics and the critical index $k_g$}
\label{sec:kg}
\begin{figure}
    \centering
\includegraphics[width=\columnwidth]{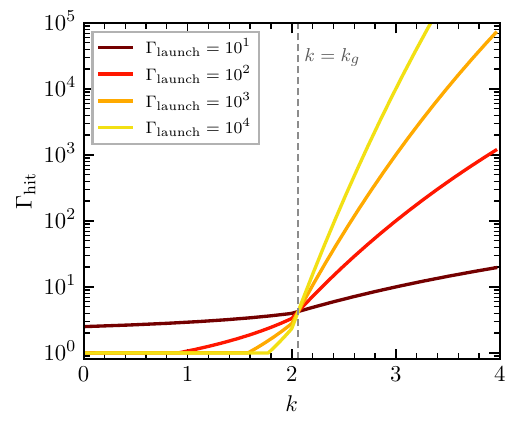}
    \caption{Shock Lorentz factor, $\Gamma_{\rm hit}$, when a forward hot-BM acoustic signal, launched from the first BM-breakdown coordinate, $\chi_b$, reaches the shock. Different curves correspond to
different launch Lorentz factors, $\Gamma_{\rm launch}$. Values for which the signal would reach the shock only after the blast wave becomes non-relativistic are plotted as $\Gamma_{\rm hit}=1$. The dashed line marks $k_g\simeq2.062$, defined by equality between the drift of the forward characteristic and that of $\chi_b$. For $k<k_g$, signals from $\chi_b$ reach the shock only after it has become trans-relativistic. For $k>k_g$, they reach it while the shock is ultra-relativistic.
}
     \label{fig:characteristic}
\end{figure}

Deep behind the shock, where the hot BM description fails, the flow is no longer determined by the BM solution but by the non-self-similar interior of the explosion. If acoustic signals emitted there could reach the shock while it is still ultra-relativistic, the BM description of the shock and of the adjacent hot shell would depend on this unknown interior. We therefore ask whether such signals can overtake the shock during the ultra-relativistic stage. Since signals propagate along the flow characteristics ($C_\pm$), we address this by following the characteristics of the hot BM solution. The characteristic speeds in the lab frame are obtained from relativistic velocity addition,
\begin{equation}
    \frac{dr_\pm}{dt}=\frac{\beta\pm c_s}{1\pm\beta c_s},
    \label{eq:vel_tran}
\end{equation}
where $c_s=\frac{1}{\sqrt{3}}$ for the hot relativistic EoS.
Using $\beta=\left(1-\frac{2}{\Gamma^2 g(\chi)}\right)^{1/2}$, one obtains, for
$\chi/\Gamma^2\ll1$ (or $\gamma^2\gg1$),
\begin{equation}
\label{eq:drdt}
    \frac{dr_\pm}{dt}
=\frac{\sqrt{1-\frac{2}{\Gamma^2g(\chi)}}\pm \frac{1}{\sqrt{3}}}{1\pm\sqrt{1-\frac{2}{\Gamma^2g(\chi)}} \cdot\frac{1}{\sqrt{3}}}=
1\pm\frac{\chi}{\Gamma^2}
\frac{1\mp\sqrt{3}}{\sqrt{3}\pm1}
+O\left(\frac{\chi^2}{\Gamma^4}\right).
\end{equation}
Transforming to the BM similarity coordinate gives\citep{best_second-type_2000} (see App. \ref{app:a} for a full derivation)
\begin{equation}
\label{eq:dchi_dt}
    \frac{d\ln\chi_\pm}{d\ln t}=(3\mp2\sqrt{3})(4-k).
\end{equation}
For $k<4$, $(3-2\sqrt{3})(4-k)<0$, so a forward-going
characteristic $C_+$ moves toward smaller $\chi$, i.e., toward the shock. A forward characteristic launched from $\chi_b$, when the shock Lorentz factor is $\Gamma_{\rm launch}$, reaches the shock at $\chi=1$, when the shock Lorentz factor is $\Gamma_{\rm hit}$,
\begin{equation}
\begin{aligned}
        \Gamma_{\rm hit}^2&=\Gamma_{\rm launch}^2\,\chi_b^\frac{3-k}{(3-2\sqrt{3})(4-k)},\\&\propto\begin{cases}
           \left(\Gamma_{\rm launch}^2\right)^{1+\frac{3-k}{(3-2\sqrt{3})(4-k)}} ,\quad &\text{for } k\leq2,\\
           \left(\Gamma_{\rm launch}^2\right)^{1+\frac{3(3-k)}{(3-2\sqrt{3})(k+4)}} ,\quad &\text{for } k>2.
           \end{cases}
           \label{eq:gamma_hit}
           \end{aligned}
\end{equation}
The critical density index $k_g$ separates two causal regimes: for
$k<k_g$, signals launched from the breakdown coordinate reach the shock
only after the blast wave has become trans-relativistic, whereas for
$k>k_g$, they reach it while the shock is still ultra-relativistic. It is defined by the condition that the forward characteristic and the
breakdown coordinate drift at equal logarithmic rates, so that a $C_+$
characteristic launched from $\chi_b$ remains on $\chi_b$,
\begin{equation}
\frac{d\ln\chi_+}{d\ln t}=\frac{d\ln\chi_b}{d\ln t}.
\end{equation}
Using the thermodynamic-transition branch of $\chi_b$, this gives
\begin{equation}
k_g=\frac{7\sqrt{3}}{2}-4\simeq2.062,
\end{equation}
which is precisely the value at which the exponent of
$\Gamma_{\rm launch}^{2}$ in Eq.~\eqref{eq:gamma_hit} vanishes. For
$k\le2$ that exponent is negative throughout: its formal zero,
$k=3-\sqrt{3}/2\simeq2.13$, the estimate obtained from $\chi_b\propto\Gamma^2$,\citep{faran_non-relativistic_2021} lies outside the $k\le2$ branch, so the causal transition occurs on the $k>2$ branch, at $k_g$. In the limit $\Gamma_{\rm launch}\to\infty$, in which the BM solution applies, the transition is sharp: for $k<k_g$ the signal reaches the shock only after the blast wave is no longer ultra-relativistic, whereas for $k>k_g$ it overtakes the shock at $\Gamma_{\rm hit}\gg1$.
At $k=k_g$, $\Gamma_{\rm hit}$ is constant, independent of
$\Gamma_{\rm launch}$.
Fig.~\ref{fig:characteristic} illustrates the behavior at finite $\Gamma_{\rm launch}$. Values
with $\Gamma_{\rm hit}<1$ are plotted as $\Gamma_{\rm hit}=1$,
corresponding to signals that arrive only after the blast wave is
non-relativistic. The curves cross at $k_g$, where
$\Gamma_{\rm hit}$ is independent of $\Gamma_{\rm launch}$. For
$k>k_g$, the hot-BM characteristic analysis is no longer sufficient,
and the cooling layer must be resolved explicitly.

The criterion generalizes to planar and cylindrical geometry
(App.~\ref{app:geometry}). There, the causal boundary is kinematic, given by a condition on the deceleration rate alone, $m=\sqrt3/2$, and
lies substantially lower, at $k_g\simeq0.134$ and $k_g\simeq1.134$, respectively. Only in the spherical case is it set by the thermodynamic transition.

\section{A cooling self-similar extension of the Blandford--McKee flow}
\label{sec:cooling}
\subsection{The cooling self-similar equations}
For \(k>2\), the hot BM solution fails thermodynamically before it fails
kinematically. The shocked fluid becomes thermodynamically trans-relativistic
($p/\rho\sim1$) while its bulk motion is still relativistic. This motivates a composite asymptotic description: an outer hot BM shell adjacent to the shock and a cooling self-similar layer behind it. To derive the cooling solution, we follow the procedure introduced by
Pan \& Sari\citep{pan_composite_2009}: we identify the point in the hot BM solution where the fluid becomes thermodynamically
trans-relativistic, define a new characteristic scale there, and derive a
new asymptotic solution around that moving transition. The resulting flow is therefore not globally self-similar, but is described by a composite of two distinct asymptotic self-similar regions.

Let \(r_{\rm cold}(t)\)
be the radius associated with
\(\chi=\chi_{\rm cold}\). We define the cooling-layer scale by
\begin{equation}
\delta(t)\equiv R-r_{\rm cold}.
\label{eq:delta_def}
\end{equation}
To leading order,
\begin{equation}
\delta\simeq
\frac{R(\chi_{\rm cold}-1)}{2(4-k)\Gamma^2}.
\label{eq:delta_scaling_hot}
\end{equation}
We introduce the cooling similarity variable
\begin{equation}
\xi\equiv \frac{R-r}{\delta},
\label{eq:xi_def}
\end{equation}
related to $\chi$ according to 
\begin{equation}
\chi=1+(\chi_{\rm cold}-1)\xi .
\label{eq:chi_xi_relation}
\end{equation}
The shock is at \(\xi=0\), the nominal hot-to-cold transition is at
\(\xi=1\), and the overlap with the hot BM solution is
\begin{equation}
\chi_{\rm cold}^{-1}\ll \xi \ll 1,
\qquad
\chi\simeq \chi_{\rm cold}\xi .
\label{eq:overlap_region}
\end{equation}
The matching condition is asymptotic: the cooling solution need not equal
the hot BM solution pointwise at \(\xi=1\). The hot BM solution fixes the scaling of the cooling branch, while the cooling profiles are determined by the self-similar equations obtained below from the scalings of the cooling layer.

To follow the fluid as it cools out of the thermally ultra-relativistic
regime, we use the following EoS,
\begin{equation}
p=\frac{\varepsilon-\rho}{3},
\qquad
\varepsilon=\rho+3p.
\label{eq:cooling_eos}
\end{equation}
This EoS describes a fluid whose adiabatic index remains $4/3$ at all
temperatures, while keeping the rest-mass term. However, the hydrodynamic equations retain their
self-similar structure for any EoS of the form $p/\rho=F(\varepsilon/\rho)$,
where $F$ is an invertible function, and the analysis below can be
generalized accordingly.\citep{pan_composite_2009}
The cooling solution is written as
\begin{equation}
\begin{aligned}
\gamma^2(r,t)&=\frac{1}{2}\overline{\Gamma}^2(t)\,
\overline{g}(\xi),\\
p(r,t)&=\overline{P}(t)\,\overline{f}(\xi),\\
\rho(r,t)&=\overline{N}(t)\,
\frac{\overline{h}(\xi)}{\sqrt{\overline{g}(\xi)}} .
\end{aligned}
\label{eq:cooling_ansatz}
\end{equation}
Here \(\overline{\Gamma}\), \(\overline{P}\), and \(\overline{N}\) are the characteristic cooling scales inherited from the hot BM solution at
\(p/\rho=1\). We choose the normalization
\begin{equation}
\overline{P}=\overline{N}.
\label{eq:Pbar_Nbar}
\end{equation}
This fixes the scaling of the new solution, not the values of
\(\overline{f}\), \(\overline{g}\), and \(\overline{h}\) at \(\xi=1\).
We define
\begin{equation}
q_\gamma\equiv \frac{d\ln\overline{\Gamma}}{d\ln t},
\qquad
q_p\equiv \frac{d\ln\overline{P}}{d\ln t}
=
\frac{d\ln\overline{N}}{d\ln t},
\qquad
q_\delta\equiv \frac{d\ln\delta}{d\ln t}.
\label{eq:q_defs}
\end{equation}
Using the BM scalings at \(\chi_{\rm cold}\), one obtains
\begin{equation}
\begin{aligned}
q_\gamma&=\frac{2(k-2)(k-3)}{k+4},\\
q_p&=\frac{4k^2-32k+39}{k+4},\\
q_\delta&=\frac{-4k^2+21k-20}{k+4}.
\end{aligned}
\label{eq:q_values}
\end{equation}
Since these exponents are constants for fixed \(k\), the leading-order hydrodynamic equations reduce to a closed system of ODEs in \(\xi\).

To reduce the hydrodynamic equations to ODEs in $\xi$, we express the
derivatives in similarity variables and identify the rate at which a
fluid element drifts through the similarity pattern,
\begin{equation}
\partial_r=-\frac{1}{\delta}\partial_\xi,
\qquad
\left.\partial_t\right|_r
=
\left.\partial_t\right|_\xi
+
\frac{\dot R-\xi\dot\delta}{\delta}\partial_\xi .
\label{eq:derivative_ops}
\end{equation}
Using
\begin{equation}
\beta=\sqrt{1-\gamma^{-2}}
\simeq
1-\frac{1}{\overline{\Gamma}^2\overline{g}}
-\frac{1}{2\overline{\Gamma}^4\overline{g}^{\,2}},
\label{eq:beta_expansion}
\end{equation}
and the Lagrangian derivative of \(\xi\),
\begin{equation}
\frac{d\xi}{dt}
=
\frac{\dot R-\beta-\xi\dot\delta}{\delta},
\label{eq:DxiDt}
\end{equation}
to the order required for the cooling-layer limit, we find
\begin{equation}
t\frac{d\xi}{dt}
=
\frac{2(4-k)}{\overline{g}}-q_\delta\xi
\equiv \lambda(\xi).
\label{eq:lambda_def}
\end{equation}
Thus \(\lambda\) is the drift velocity of a fluid element through the
cooling similarity pattern.

The similarity equations are derived from the conservative spherical
relativistic hydrodynamic equations, Eqs.~(\ref{eq:relativistic_euler}). The reduction is carried out in the asymptotic regime $\overline{\Gamma}\gg1,
\gamma\gg1$. One must retain subleading terms before differentiating combinations
that are degenerate at leading order. In particular, the equation
obtained from energy minus momentum requires the first
\(O(\overline{\Gamma}^{-2})\) correction before the radial derivative is
taken.

The resulting self-similar ODE system is
\begin{equation}
\begin{aligned}
\left(\frac{2(4-k)}{\overline{g}}-q_\delta\xi\right)
\frac{\overline{h}'}{\overline{h}}
-2(4-k)\frac{\overline{g}'}{\overline{g}^{\,2}}
+(q_\gamma+q_p+2)
&=0,
\\
-q_\delta\xi\left(
\overline{f}'
+\frac{\overline{h}'}{2\sqrt{\overline{g}}}
-\frac{\overline{h}\,\overline{g}'}
{4\overline{g}^{3/2}}
\right)
+(q_p+4)\overline{f}
+\frac{q_p+2}{2}
\frac{\overline{h}}{\sqrt{\overline{g}}}
&\\
+4(4-k)\left(
\frac{\overline{f}'}{\overline{g}}
-\frac{\overline{f}\,\overline{g}'}
{\overline{g}^{\,2}}
+\frac{\overline{h}'}{4\overline{g}^{3/2}}
-\frac{3\overline{h}\,\overline{g}'}
{8\overline{g}^{5/2}}
\right)
&=0,
\\
(2q_\gamma+q_p+2)
\left(
2\overline{f}\,\overline{g}
+\frac{\overline{h}\sqrt{\overline{g}}}{2}
\right)
&\\
-q_\delta\xi\left(
2\overline{g}\,\overline{f}'
+2\overline{f}\,\overline{g}'
+\frac{\sqrt{\overline{g}}}{2}\overline{h}'
+\frac{\overline{h}\,\overline{g}'}
{4\sqrt{\overline{g}}}
\right)
&\\
+2(4-k)\left(
\overline{f}'
+\frac{\overline{h}'}{2\sqrt{\overline{g}}}
-\frac{\overline{h}\,\overline{g}'}
{4\overline{g}^{3/2}}
\right)
&=0,
\end{aligned}
\label{eq:ode_cooling}
\end{equation}
 where primes ($'$) denote derivatives with respect to \(\xi\). The cooling branch must approach the hot BM solution in the overlap region, so for
\(\xi\rightarrow0\) ($\xi\propto\chi$, see Eq.~(\ref{eq:overlap_region})) the solution approaches
\begin{equation}
\overline{g}(\xi)\sim \xi^{-1},
\qquad
\overline{f}(\xi)\sim \xi^{\alpha_f},
\qquad
\overline{h}(\xi)\sim \xi^{\alpha_h},
\label{eq:hot_overlap}
\end{equation}
where
\begin{equation}
\alpha_f=\frac{4k-17}{12-3k},
\qquad
\alpha_h=\frac{2k-7}{4-k}.
\label{eq:BM_overlap_exponents}
\end{equation}
Eqs.~\eqref{eq:ode_cooling} admit this
BM overlap behavior.

\subsection{Characteristic structure and the sonic line}
\label{sec:sonic_regular}
Before constructing the cooling profiles, we map the causal structure of the solution: it determines both where the BM-connected branch
terminates and, as shown in \S~\ref{sec:selection}, how the physical solution is selected.
The characteristic structure of the cooling solution follows from the motion of acoustic signals in the cooling similarity coordinate. A thin shell of constant \(\xi\) has radius \(r_\xi(t)=R(t)-\xi\delta(t)\), and therefore velocity \(\dot r_\xi=\dot R-\xi\dot\delta\). In terms of the similarity coordinate, we find 
\begin{equation}
\frac{d\xi_+}{dt}
=
\frac{\dot r_\xi-\dot r_+}{\delta}.
\label{eq:dxiplus_general}
\end{equation}
In the proper asymptotic limit, we have (see App. \ref{app:b} for full derivation)
\begin{equation}
t\frac{d\xi_+}{dt}
=
\frac{2(4-k)}{\overline{g}}
\left[
\frac{1-c_s}{1+c_s}-z
\right],
\qquad
z\equiv \frac{q_\delta\xi\overline{g}}{2(4-k)},
\label{eq:dxiplus}
\end{equation}
where $c_s$ is the local sound speed,
\begin{equation}
c_s^2\equiv\left(\frac{\partial p}{\partial\varepsilon}\right)_{\!s}
     =\frac{4\eta}{3(1+4\eta)},
\qquad
\eta\equiv\frac{\overline{f}\sqrt{\overline{g}}}{\overline{h}} .
\label{eq:cs2}
\end{equation}
We define
\begin{equation}
\mathcal{D}_+
\equiv
\frac{1-c_s}{1+c_s}-z,
\label{eq:Dplus_zero}
\end{equation}
so that a forward acoustic characteristic is stationary in the similarity
coordinate, where $\mathcal{D}_+=0$. Such a point will be referred to as a sonic point.
In the hot overlap, $\eta\to\infty$, $c_s\to1/\sqrt3$, and
$z\to z_{\rm hot}=(4k-5)/[2(k+4)]$, so that
$\mathcal{D}_+\to2-\sqrt3-(4k-5)/[2(k+4)]$, which changes sign at $k_g\simeq2.062$, reproducing the hot-BM characteristic criterion of \S \ref{sec:kg}. For $k>k_g$ the outer cooling layer is shock connected, $\mathcal{D}_+<0$, and the causal structure deeper in the flow is set by the sonic point reached along the solution.

To locate the sonic points of the solution, it is convenient to work with the logarithmic derivatives
\begin{equation}
\overline{f}_\xi\equiv \frac{d\ln \overline{f}}{d\ln \xi},
\qquad
\overline{g}_\xi\equiv \frac{d\ln \overline{g}}{d\ln \xi},
\qquad
\overline{h}_\xi\equiv \frac{d\ln \overline{h}}{d\ln \xi}.
\label{eq:uvw_def}
\end{equation}
In these variables, the continuity equation is
\begin{equation}
\overline{g}_\xi-(1-z)\overline{h}_\xi=\frac{(q_\gamma+q_p+2)z}{q_\delta} ,
\label{eq:cont_slope}
\end{equation}
and elimination of $\overline{h}_\xi$ from the remaining two equations yields the linear
system
\begin{equation}
\begin{aligned}
(2-z)\,\overline{f}_\xi+\left(\frac{z-1}{4\eta}-2\right)\overline{g}_\xi
 &=\frac{z}{q_\delta}\left[\frac{q_\gamma}{2\eta}-(q_p+4)\right],
\\
(1-2z)\,\overline{f}_\xi+\left(\frac{1-z}{4\eta}-2z\right)\overline{g}_\xi
 &=\frac{z}{q_\delta}\left[-2(2q_\gamma+q_p+2)-\frac{q_\gamma}{2\eta}\right].
\label{eq:uv}
\end{aligned}
\end{equation}
Since $\overline{f}_\xi$, $\overline{g}_\xi$ and $\overline{h}_\xi$ are determined algebraically by $(z,\eta)$, and
since
\begin{equation}
\frac{d\ln z}{d\ln\xi}=1+\overline{g}_\xi,\qquad
\frac{d\ln\eta}{d\ln\xi}=\overline{f}_\xi+\frac{\overline{g}_\xi}{2}-\overline{h}_\xi,
\label{eq:autonomous} 
\end{equation} 
the similarity equations reduce to an autonomous flow in the $(z,\eta)$ plane. The hot BM overlap is the fixed point
$(z,\eta)=(z_{\rm hot},\infty)$ of this flow, and the cooling branch is the
trajectory emanating from it. This reduction organizes the construction of the solution and its sonic points, to which we now turn.

\begin{figure} 
\centering \includegraphics[width=\columnwidth]{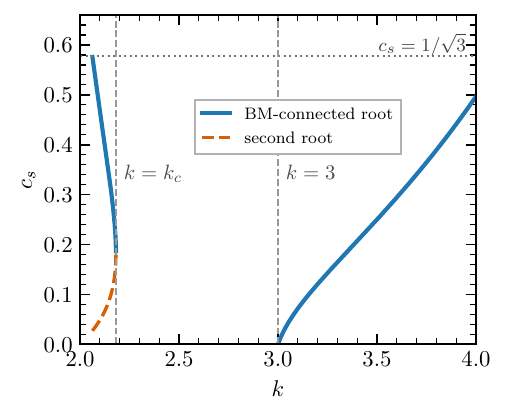} \caption{Physical roots of the finite-sound-speed regularity condition, Eq.~\eqref{eq:csquad}, i.e., only roots satisfying \(0\leq c_s\leq1/\sqrt{3}\) are shown. For \(k_g<k<k_c\) there are two roots: the BM-connected endpoint is the larger (upper) one, which runs from \(c_s=1/\sqrt{3}\) at \(k_g\simeq2.062\) down to \(c_s\simeq0.183\) at \(k_c\simeq2.181\), where the two roots merge. A second finite-\(c_s\) branch exists for \(k>3\), with \(c_s\rightarrow0\) as \(k\rightarrow3^+\). In the intermediate range \(k_c< k\le3\), no positive finite-\(c_s\) endpoint is available, and the BM-connected branch reaches the forward sonic condition in the zero-pressure limit.} 
\label{fig:finite_cs_roots}
\end{figure}

The determinant of the system
(Eq.~\eqref{eq:uv}) is proportional to
$3(1-z)^2+8\eta\,(z^2-4z+1)$ and vanishes precisely on the sonic line, $\mathcal{D}_+=0$. A smooth crossing, therefore, requires the corresponding numerator to
vanish as well, which leads to the following regularity condition
\begin{equation}
3q_p c_s^2+\left(4q_\gamma+3q_p+8\right)c_s+4q_\gamma=0 .
\label{eq:csquad}
\end{equation}
Physical roots are those with $0\leq c_s\leq1/\sqrt3$. Their structure is shown
in Fig.~\ref{fig:finite_cs_roots}.
A pair of physical roots exists for
$k_g<k<k_c$, where
\begin{equation}
k_c\simeq2.181
\label{eq:kc} 
\end{equation}
is the smaller positive value of $k$ at which the discriminant vanishes,
$\left(4q_\gamma+3q_p+8\right)^2=48\,q_p q_\gamma$, equivalently
$16k^4-128k^3+2856k^2-13280k+16345=0$. The two roots merge at $k_c$ with $c_s\simeq0.1834$ and become complex beyond it. They reappear as real
but negative (unphysical) roots for $2.976\lesssim k\le3$, and a single physical root exists for $3<k<4$, with $c_s\to0^+$ as $k\to3^+$.

Two roots are available on $k_g<k<k_c$, but only the larger one is the
endpoint of the BM-connected branch. It is the root that is
continuous with the overlap criterion: as $k\to k_g^+$, the larger root
approaches the hot value $c_s=1/\sqrt3$, and the sonic point recedes into
the hot overlap itself. Since the gas cools monotonically inward, the sound speed falls steadily from $1/\sqrt{3}$, so this larger $c_s$ point is the one encountered first. It is the inner causal boundary, the sonic horizon beyond which downstream acoustic characteristics can no longer overtake the shock. The smaller $c_s$ root lies deeper in and plays no physical role.

When Eq.~\eqref{eq:csquad} has no physical root, for $k_c<k\le3$, the branch cannot cross the sonic line smoothly at finite sound speed. Instead, it reaches a sonic point in the zero-pressure limit,
\begin{equation}
\overline{f}\to0,\qquad c_s\to0,\qquad z\to1,\qquad \lambda\to0 ,
\label{eq:cornerlimit}
\end{equation}
where $\lambda$ is the drift velocity (Eq.~\eqref{eq:lambda_def}).
This is the corner $(z,\eta)\to(1,0)$ of the reduced plane, which is the
$c_s\to0$ limit of the sonic line $\mathcal{D}_+=0$. Although the
logarithmic slope $\overline{f}_\xi$ diverges there, the combination
\begin{equation}
\eta \overline{f}_\xi
=\frac{\xi\overline{f}'\sqrt{\overline{g}}}{\overline{h}}
\label{eq:Pdef}
\end{equation}
remains finite, and evaluating Eqs.~(\ref{eq:ode_cooling}) in this limit gives
the finite pressure-gradient condition (see App. \ref{app:corner} for full derivation),
\begin{equation}
\left.\frac{\xi\overline{f}'\sqrt{\overline{g}}}{\overline{h}}\right|_{\xi_s}
 =\frac{q_\gamma}{2q_\delta} .
\label{eq:corner_grad}
\end{equation}
This condition replaces the finite $c_s$
regularity condition at the zero-pressure limit: while Eq.~\eqref{eq:csquad} fixes the sonic point sound speed, Eq.~\eqref{eq:corner_grad} fixes the sonic pressure gradient. At a finite-$c_s$ sonic point, the regularity condition Eq.~\eqref{eq:csquad} makes all profile slopes finite, so the density is finite there. At the zero-pressure endpoint, by contrast, Eq.~\eqref{eq:corner_grad} keeps only the scaled pressure gradient finite. The pressure vanishes, and the Lorentz factor remains finite, but the density does not in general remain finite. Solving the endpoint equations together with the approach to $(z,\eta)\to(1,0)$ (see App. \ref{app:corner} for full derivation) gives a power-law density profile,
\begin{equation}
\rho\propto(\xi_s-\xi)^{\mu_\rho},\qquad \mu_\rho=\frac{12-5k}{k},
\label{eq:rho_power}
\end{equation}
with the lab-frame density $\gamma\rho\propto\overline{h}$ carrying the same exponent, since $\overline{g}$ is finite at the endpoint. The exponent changes sign at $k=12/5$. The density, therefore, vanishes for $k_c<k<12/5$, finite only at $k=12/5$, and diverges for $12/5<k<3$. For $k_c<k<12/5$, the zero-pressure endpoint is an evacuated cavity (hollow solution). For $12/5<k<3$ the density diverges at vanishing pressure. The divergence is nonetheless integrable, since $\mu_\rho>-1$ for all $k<3$ (it reaches $-1$ only in the degenerate limit $k\to3$), so the enclosed mass $\propto\int(\xi_s-\xi)^{\mu_\rho}d\xi$ remains finite. An analogous endpoint structure is known in the Newtonian
strong-explosion problem, where for $2<\omega<3$ the solutions
terminate at a zero-pressure point with an integrable density
singularity, enclosing an evacuated
interior.\citep{sedov_similarity_1959,gruzinov_self-similarity_2003} The self-similar description, therefore, simply terminates at $\xi_s$ and must be matched there to the unknown interior. This inner continuation may involve additional structure, such as an inner shock\citep{faran_non-relativistic_2021} or a contact discontinuity, but it is not determined by, and not required for, the present construction. The causal shielding of the hot shell is unaffected in either case, since the endpoint remains a forward causal boundary beyond which downstream acoustic characteristics cannot overtake the shock.

The endpoint regimes are summarized in Table~\ref{tab:regimes}. Within the
cold regime $k_c<k<3$ the endpoint density follows
Eq.~\eqref{eq:rho_power}: it vanishes for $k_c<k<12/5$, where the solution is hollow, and diverges (but integrable) for $12/5<k<3$.
\begin{table}
\caption{Inner sonic point of the BM-connected cooling flow for each range
of the external density index $k$. The pressure is
finite at the sonic point in the two finite-$c_s$ regimes and vanishes at
the cold sonic point. In the cold regime, the endpoint density follows
Eq.~\eqref{eq:rho_power}: it vanishes (hollow) for $k_c<k<12/5$ and
diverges (but integrable) for $12/5<k<3$.}
\label{tab:regimes}
\centering
\begin{tabular}{cc}
\hline\hline
$k$ & inner sonic point \\
\hline
$k_g<k<k_c$ & finite-$c_s$, finite $\rho$\\
$k_c<k<12/5$ & cold $c_s=0$, hollow ($\rho\to0$)\\
$12/5<k<3$   & cold $c_s=0$, $\rho\to\infty$ (integrable)\\
$3<k<4$     & finite-$c_s$, finite $\rho$\\
\hline\hline
\end{tabular}
\end{table}

\subsection{Selection of the BM-connected branch}
\label{sec:selection}
The inner sonic point identified in \S\ref{sec:sonic_regular} fixes where the shock-connected cooling flow terminates, but not which cooling solution reaches it. Matching the hot BM shell leaves a one-parameter family. To see this, we linearize the flow (Eq.~\eqref{eq:autonomous}) about the overlap fixed point $(z,\eta)=(z_{\rm hot},\infty)$, where the flow
coincides with the hot BM solution. The solution near the overlap is the hot BM profile plus the corrections introduced by the softening of the EoS as the gas cools, which decay toward the shock as
\begin{equation}
\eta^{-1}\propto\xi^{\lambda_\eta},\qquad
\lambda_\eta=\frac{k+4}{6(4-k)} ,
\label{eq:lameta}
\end{equation}
along with a single free mode,
\begin{equation}
\overline{g}\xi-1\propto C\,\xi^{\lambda_z},\qquad
\lambda_z=\frac{(k+4)\left(59+4k-4k^2\right)}{3(4-k)\left(4k^2+32k-83\right)},
\label{eq:lamz}
\end{equation}
whose amplitude $C$ the matching leaves undetermined. Since $\lambda_z>0$
throughout $k_g<k<4$, this mode also decays toward the shock and leaves no
trace in the overlap. The denominator of Eq.~\eqref{eq:lamz} factorizes
as $4k^2+32k-83=4(k-k_g)(k+4+7\sqrt3/2)$, so $\lambda_z\to\infty$ as
$k\to k_g^+$ and the mode decouples from the shock infinitely
steeply at the causal boundary $k_g$. The amplitude $C$ is thus the single
free parameter of the problem, fixed not by the shock but by the
requirement that the flow reach the inner sonic point cleanly.
The physical solution is thus the one value of $C$ whose trajectory terminates
regularly at the sonic point. In the $(z,\eta)$ plane, the selected solution is the single trajectory that leaves the hot overlap and reaches the sonic point identified in \S~\ref{sec:sonic_regular}.

\begin{figure}
\centering
\includegraphics[width=\columnwidth]{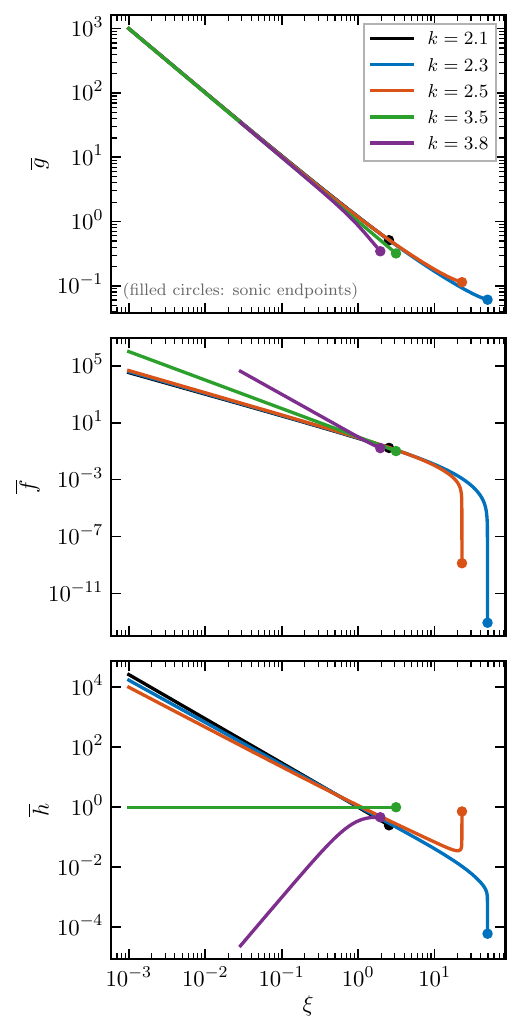}
\caption{Representative BM-connected cooling profiles for five external
density indices, $k=2.1,\,2.3,\,2.5,\,3.5,\,3.8$ (colors), spanning the
finite-$c_s$ and cold endpoint regimes. Panels show $\overline{g}$, $\overline{f}$, $\overline{h}$
(Eq.~\eqref{eq:cooling_ansatz}) versus $\xi$, where each
profile approaches its hot-BM overlap form, $\overline{g}\sim\xi^{-1}$, $\overline{f}\sim\xi^{\alpha_f}$,
$\overline{h}\sim\xi^{\alpha_h}$ as $\xi\to0$. Filled circles mark the
inner sonic endpoint $\xi_s(k)$. At a finite-$c_s$ endpoint, the pressure
stays finite. At a cold endpoint $\overline{f}\to0$, while the density
vanishes (hollow, $k_c<k<12/5$) or diverges ($12/5<k<3$),
following Eq.~\eqref{eq:rho_power}.}
\label{fig:cooling_profiles}
\end{figure}

\begin{figure}
\centering
\includegraphics[width=\columnwidth]{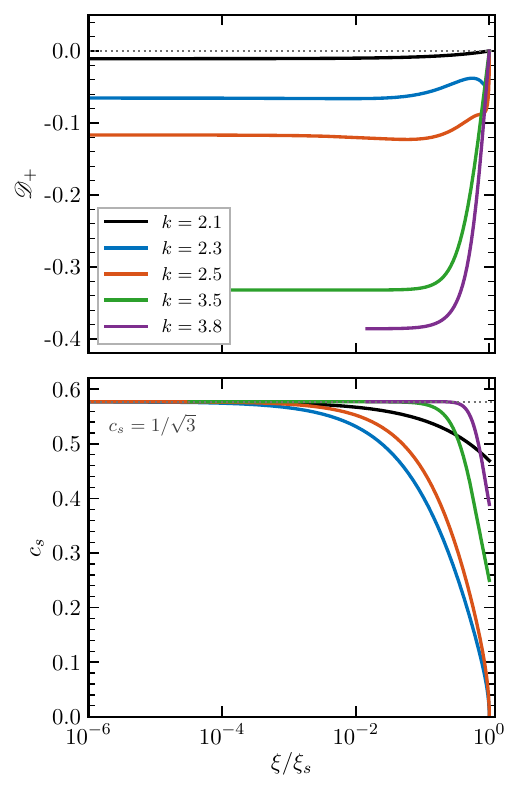}
\caption{Characteristic diagnostics along the same five
BM-connected cooling solutions as in Fig.~\ref{fig:cooling_profiles}, plotted versus $\xi/\xi_s$. Top: the
forward-characteristic function, $\mathcal{D}_+$
(Eq.~\eqref{eq:Dplus_zero}), stays negative throughout the cooling
layer (shock-connected) and rises to
$\mathcal{D}_+=0$ (dotted line) at the sonic point. Bottom: the local
sound speed $c_s$. In the hot overlap ($\xi/\xi_s\to0$) every branch
approaches the relativistic value, $c_s=1/\sqrt{3}$ (dotted line), and then
falls steadily as the gas cools inward, ending at $c_s>0$ (finite-$c_s$ branches)
or $c_s\to0$ (cold branches).}
\label{fig:critical_diagnostics}
\end{figure}

Because both modes, Eqs.~\eqref{eq:lameta}--\eqref{eq:lamz}, grow inward, the solution cannot be built by integrating inward from the overlap, where any small error is
amplified by many orders of magnitude across the layer. We instead construct the branch as a boundary-value problem, numerically integrating outward from the sonic point toward the shock, the direction in which both corrections decay, and the hot BM solution is recovered. In this direction, the overlap fixed point is attracting, and the
trajectory reaching it from the sonic point is unique in the
$(z,\eta)$ plane, so $C$ is
obtained as an output of a single integration. At a finite-$c_s$ endpoint, the
slopes are of indeterminate ($0/0$) form, since the regularity condition,
Eq.~\eqref{eq:csquad}, makes the numerator of the solution to Eq.~\eqref{eq:uv} vanish together with the determinant, following the standard sonic-point construction of second-type self-similar solutions. \citep{waxman_secondtype_1993,sari_first_2006} The integration is therefore launched along the direction of the linearized equations that connects to the overlap. At a zero-pressure endpoint, no such indeterminacy arises: the determinant of
Eq.~\eqref{eq:uv} stays finite there, and the solution approaches the corner
along the analytic direction $\eta/(1-z)\to(k-2)/(13-2k)$ (App.~\ref{app:corner}), from which the integration
is launched directly. In either case, the
overlap behavior, Eq.~\eqref{eq:hot_overlap}, imposed at unit amplitude, then
leaves no freedom, so the location of the sonic point $\xi_s(k)$ comes out
of the construction rather than being imposed. Fig.~\ref{fig:cooling_profiles}
shows the resulting profiles, and Fig.~\ref{fig:critical_diagnostics} the
corresponding characteristic diagnostics.

\begin{figure}
\centering
\includegraphics[width=\columnwidth]{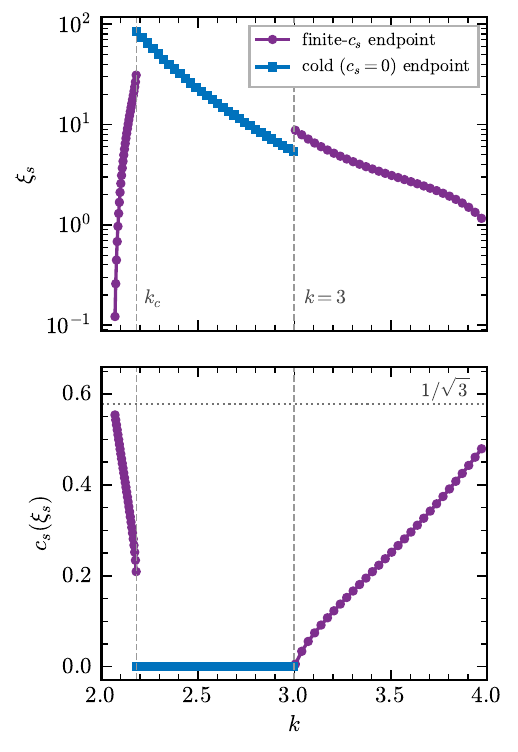}
\caption{Location of the inner sonic point, $\xi_s$, and the corresponding sound speed, $c_s(\xi_s)$, of the BM-connected cooling flow
as a function of $k$. Purple points are finite-sound-speed sonic points, blue
points are the cold (zero-pressure) sonic points. The location $\xi_s(k)$ is an
output of the construction, continuous within each regime, and jumping at
the regime boundaries, $k_c$ and $k=3$, each jump reflecting a change in
the type of sonic point reached.}
\label{fig:endpoint_map}
\end{figure}

The location of the inner sonic point and its corresponding sound speed as a function of $k$, are given in Fig.~\ref{fig:endpoint_map}. The inner boundary, $\xi_s(k)$, jumps at $k_c$ and at $k=3$. For example, approaching $k_c$ from below, the two finite-$c_s$ sonic points merge, with $c_s\simeq0.183$, and the sonic point sits at $\xi_s\simeq36$, while just above $k_c$ the flow reaches the cold sonic point much deeper in, at $\xi_s\simeq85$. The jump is a change in the kind of sonic point the flow reaches, and on each side, the solution is unique. At $k_c$ the flow itself varies continuously in $k$, and only the location of the first sonic crossing, the causal boundary, jumps. At $k=3$ the pressure-gradient condition (Eq.~\eqref{eq:corner_grad}) degenerates ($q_\gamma=0$), the approach ratio is left free (App.~\ref{app:corner}), and the two
branches attach to solutions with different ratios, at $\xi_s\simeq5.3$ and $\xi_s\simeq8.9$. The deep part of the cooling layer is therefore discontinuous at $k=3$, while the difference decays toward the shock, so the near-shock flow varies continuously in $k$.

\subsection{Analytic solution at \(k=7/2\)}
\label{sec:k72_exact}
For a general external density index \(k\), the cooling solution must be constructed numerically. A useful exception occurs at
\(k=7/2\), for which the similarity equations admit a closed-form analytic
solution. At this value, the BM-overlap exponents are $\alpha_f=-2, \alpha_h=0.$ The overlap scalings are therefore $\overline{g}\sim \xi^{-1},  \overline{f}\sim \xi^{-2}, \overline{h}\sim 1.$
Substitution into the cooling equations shows that, for \(k=7/2\), these scalings are not merely asymptotic: they solve Eqs.~(\ref{eq:ode_cooling})
exactly,
\begin{equation}
\overline{g}(\xi)=\xi^{-1},
\qquad
\overline{f}(\xi)=\xi^{-2},
\qquad
\overline{h}(\xi)=1 .
\label{eq:k72_exact_solution}
\end{equation}
Thus, the cooling solution is known in closed form.

The characteristic structure of this exact branch is also explicit. For
\(k=7/2\), we find
\begin{equation}
q_\delta=\frac{3}{5}\rightarrow\quad z\equiv \frac{q_\delta \xi \overline{g}}{2(4-k)}
=
\frac{(3/5)\xi\,\xi^{-1}}{2(1/2)}
=
\frac{3}{5},
\end{equation}
so \(z\) is constant along the solution. The sound speed on the exact
branch is
\begin{equation}
c_s^2
=
\frac{4\overline{f}\sqrt{\overline{g}}}
{3\left(\overline{h}
+4\overline{f}\sqrt{\overline{g}}\right)}
=
\frac{4\xi^{-5/2}}
{3\left(1+4\xi^{-5/2}\right)}.
\label{eq:csxi}
\end{equation}
Since \(z=3/5\), the intersection with the sonic point occurs when
\begin{equation}
\frac{1-c_s}{1+c_s}=\frac{3}{5}\rightarrow\quad c_s=\frac{1}{4}.
\end{equation}
Substituting this value into Eq.~\eqref{eq:csxi} gives
\begin{equation}
\xi_s=\left(\frac{52}{3}\right)^{2/5}.
\end{equation}

The \(k=7/2\) solution provides an explicit closed-form example of the
BM-connected cooling branch. It approaches the hot BM scalings as
\(\xi\rightarrow0\), evolves through the cooling layer, and reaches a
finite-sound-speed sonic point. The corresponding characteristic
line in the physical \((r,t)\) plane marks the inner boundary of the
BM-connected composite solution. This analytic solution, therefore,
provides a benchmark for the numerical critical-point construction and
makes the causal shielding mechanism explicit.

\begin{figure}
    \centering
    \includegraphics[width=\columnwidth]{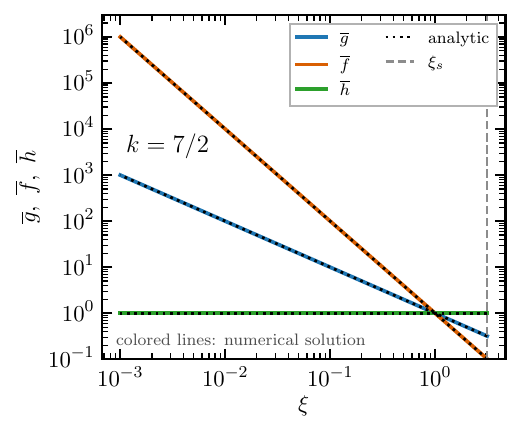}
    \caption{BM-connected cooling solution for \(k=7/2\). The analytic solution is \(\overline{g}=\xi^{-1}\), \(\overline{f}=\xi^{-2}\), and \(\overline{h}=1\). Along this branch \(z=3/5\), and the forward sonic point is reached at \(c_s=1/4\), corresponding to \(\xi_s=(52/3)^{2/5}\). The numerical construction (colored solid lines) reproduces the analytic profile (dotted black line) and critical point.} \label{fig:k72_exact}
\end{figure}

Fig.~\ref{fig:k72_exact} compares the numerical critical-branch
construction with the analytic \(k=7/2\) solution. The agreement verifies
both the normalization to the BM overlap and the identification of the
sonic point.

\subsection{Domain of validity}
\label{sec:cooling_validity}
The cooling solution derived above is an asymptotic layer attached to the
hot BM shell. Its domain of validity is therefore limited by the
assumptions used in its construction: the shock and the cooling layer
bulk motion must remain ultra-relativistic, the cooling layer must remain
thin compared with the shock radius, and an overlap region with the hot
BM solution must exist (Eq.~(\ref{eq:overlap_region})). Within this domain, the cooling solution should not be interpreted as a
global continuation of the downstream flow to arbitrary depth. It is the
part of the post-BM flow selected by asymptotic matching to the hot shell
and by regularity at the sonic point.

This distinction is important. The branch obtained here is the
BM-connected branch: it describes the outer cooling layer that can
communicate with the shock and that supplies the continuation of the hot
BM shell after the thermodynamic transition. The branch terminates at the
sonic point discussed above. The flow deeper downstream may require an additional inner boundary condition.

For large \(k\), especially close to \(k=4\), the sonic point construction remains the same, but the overlap region becomes harder to
realize in finite examples. Since
\begin{equation}
\chi_{\rm cold}\propto \Gamma^{6(4-k)/(k+4)},
\end{equation}
the exponent becomes small as \(k\rightarrow4\). Even very large shock
Lorentz factors then produce only a modest separation between the shock
and the cooling transition. In this limit, the cooling solution should be understood as the leading asymptotic branch obtained after taking \(\Gamma\rightarrow\infty\), rather than as a quantitatively accurate description.

The zero-sound-speed endpoint should also be interpreted within the
adopted EoS, \(\varepsilon=\rho+3p\). This EoS
includes rest-mass inertia, but it does not transition to the
non-relativistic thermal relation \(\varepsilon=\rho+3p/2\). A more complete
EoS may modify the detailed inner continuation of the zero-sound-speed branch.

\subsection{Causal shielding of the hot BM shell}
\label{sec:causal}
What the cooling solution establishes is most clearly stated in causal terms. The hot BM shell need not be causally disconnected from the entire
cooling layer. On the contrary, for $k_g<k<4$, the outer cooling layer remains connected to the shock and is matched asymptotically to the hot shell, so the two form a single composite solution. The key point is that this BM-connected composite has an inner characteristic boundary at the sonic point of
\S~\ref{sec:sonic_regular} (finite- or zero-sound-speed, depending on \(k\)). Beyond it, forward acoustic characteristics no longer propagate
toward the shock in the cooling similarity coordinate, so signals from the deeper, uncontrolled interior cannot reach the shock or the hot shell while the blast wave remains ultra-relativistic. This is the sense in which the hot BM shell is causally shielded.

\section{Summary and discussion}
\label{sec:summary}
We have studied whether the BM self-similar solution remains self-consistent despite the deeper downstream region where its assumptions fail. The BM solution is an asymptotic description of a spherical ultra-relativistic blast wave and of the hot shell immediately behind it. It is not expected to describe the downstream flow arbitrarily far behind the shock. Its breakdown deep in the flow is therefore not, by itself, a failure of the shock-local solution. The relevant question is whether the region where the BM assumptions first fail can imprint on the shock and the hot shell while the blast wave is still ultra-relativistic.

The hot BM branch can fail in two ways: kinematically, when the downstream
flow becomes only trans-relativistic, or thermodynamically, when the
shocked fluid leaves the hot EoS. For \(k<2\) the kinematic
breakdown occurs first, while for \(k>2\) the thermodynamic cooling transition occurs first, with the bulk flow still relativistic. Comparing
the drift of the breakdown coordinate with that of the forward hot-BM
characteristics gives the critical index \(k_g=7\sqrt{3}/2-4\simeq2.062\).
For \(k<k_g\), the characteristic structure of the hot BM solution alone
shields the shell, since signals launched from the region where the hot BM description first fails
reach the shock only after the blast wave has become non-relativistic. For
\(k>k_g\), this argument is insufficient, and we resolved the cooling layer explicitly by constructing a second self-similar solution that solves the relativistic hydrodynamic equations with a trans-relativistic EoS and is matched asymptotically to the hot BM shell.

This BM-connected cooling solution is selected by regularity at its inner sonic point, together with the matching to the hot overlap. The sonic point
is reached at finite sound speed for \(k_g<k<k_c\) and for \(3<k<4\), and
in the cold limit \(c_s\to0\) for \(k_c<k<3\), with \(k_c\simeq2.181\). In
the cold regime, the endpoint density follows
$\rho\propto(\xi_s-\xi)^{(12-5k)/k}$, so it vanishes for $k_c<k<12/5$ and
diverges, with finite enclosed mass, for $12/5<k<3$. In the latter range, the self-similar branch terminates at the sonic point and is matched to an unknown inner structure, and the causal shielding holds in either case. In
the physical \((r,t)\) plane, this sonic point is the inner boundary of the BM-connected composite. The hot BM shell and the cooling layer behind it
are causally connected to the shock and form a single asymptotic solution,
while the deeper region beyond the sonic point, which this construction
does not determine, cannot send acoustic information back to the
shock while the blast wave is ultra-relativistic. The BM shell, therefore, remains locally self-consistent even though the full downstream flow is not
globally described by the hot BM solution. The analytically solvable case
\(k=7/2\), with \(\overline{g}=\xi^{-1}\), \(\overline{f}=\xi^{-2}\),
\(\overline{h}=1\) and a sonic point at \(c_s=1/4\),
\(\xi_s=(52/3)^{2/5}\), realizes this picture in closed form.

The main conclusion is that the causal structure is what enables the BM solution near the shock. The BM similarity profiles are determined one-sidedly, from the shock jump conditions alone, with no condition imposed at the inner end. This construction is legitimate because the region where the BM assumptions fail cannot imprint on the shell. For $k<k_g$, signals from that region reach the shock only after the blast wave is no longer ultra-relativistic. For $k_g<k<4$, the part of it that communicates with the shock is the cooling layer, which is fixed by the shock conditions together with regularity at its inner sonic point, and its corrections decay toward the shock, while the flow beyond the sonic point, which the construction does not determine, cannot act on the shell at all. The sonic point thus supplies the missing inner boundary. The shock scaling itself is not at risk: it is anchored by energy conservation, and an interior of negligible energy cannot alter it. What a causally connected undetermined interior can do is imprint on the profiles behind the shock, so that the flow near the shock would depend on the unknown interior. Energy conservation thus fixes the exponent, while the causal boundary is what makes the profiles self-consistent. The distinction is sharpest for $k>3$: there, the deep interior dominates the swept-up rest mass so it retains the memory of the initial conditions, and the sonic point places it entirely beyond reach.

The causal criterion extends to planar and cylindrical geometry (App.~\ref{app:geometry}). On the kinematic branch, the critical condition is a condition on the deceleration rate alone, $m=\sqrt3/2$, so the critical index shifts down by one unit of $k$ per dimension, with $k_g\simeq0.134$ in planar and $k_g\simeq1.134$ in cylindrical geometry. Only in the spherical case does the thermodynamic transition set the boundary, and the causal protection established in this paper is therefore increasingly fragile in lower-dimensional geometry. For $k>k_g$, the analysis requires the general-geometry counterparts of the cooling layer and of the trans-relativistic interior,\citep{faran_non-relativistic_2021} and we defer this, together with the implications for the planar first-type solutions,\citep{faran_self-similarity_2024} to future work.

\begin{acknowledgments}
This work was in part supported by the Alexander von Humboldt Foundation. DK is supported by a research grant from The Abramson Family Center for Young Scientists, an ISF grant, the Minerva Stiftung, and the Pazi Foundation.
\end{acknowledgments}

\section*{Data Availability Statement}
The data that support the findings of this study are available from the corresponding author upon reasonable request.

\bibliography{references}

@article{granot_off-axis_2018,
	title = {Off-axis emission of short {GRB} jets from double neutron star mergers and {GRB} {170817A}},
	volume = {481},
	issn = {0035-8711},
	url = {https://ui.adsabs.harvard.edu/abs/2018MNRAS.481.1597G},
	doi = {10.1093/mnras/sty2308},
	urldate = {2026-08-31},
	journal = {Monthly Notices of the Royal Astronomical Society},
	publisher = {OUP},
	author = {Granot, Jonathan and Gill, Ramandeep and Guetta, Dafne and De Colle, Fabio},
	month = dec,
	year = {2018},
	note = {ADS Bibcode: 2018MNRAS.481.1597G},
	pages = {1597--1608},
}

@article{sari_spectra_1998,
	title = {Spectra and {Light} {Curves} of {Gamma}-{Ray} {Burst} {Afterglows}},
	volume = {497},
	issn = {0004-637X},
	url = {https://iopscience.iop.org/article/10.1086/311269/meta},
	doi = {10.1086/311269},
	number = {1},
	urldate = {2022-09-20},
	journal = {The Astrophysical Journal},
	publisher = {IOP Publishing},
	author = {Sari, Re'em and Piran, Tsvi and Narayan, Ramesh},
	month = mar,
	year = {1998},
	pages = {L17},
}

@article{faran_self-similarity_2024,
	title = {Self-similarity of the third type in ultra-relativistic blastwaves},
	volume = {36},
	issn = {0899-82131070-6631},
	url = {https://ui.adsabs.harvard.edu/abs/2024PhFl...36e6119F},
	doi = {10.1063/5.0203812},
	urldate = {2025-03-10},
	journal = {Physics of Fluids},
	publisher = {AIP},
	author = {Faran, Tamar and Gruzinov, Andrei and Sari, Re'em},
	month = may,
	year = {2024},
	pages = {056119},
}

@misc{nayana_most_2025,
	title = {The {Most} {Luminous} {Known} {Fast} {Blue} {Optical} {Transient} {AT} 2024wpp: {Unprecedented} {Evolution} and {Properties} in the {X}-rays and {Radio}},
	shorttitle = {The {Most} {Luminous} {Known} {Fast} {Blue} {Optical} {Transient} {AT} 2024wpp},
	url = {https://ui.adsabs.harvard.edu/abs/2025arXiv250900952N},
	doi = {10.48550/arXiv.2509.00952},
	urldate = {2025-09-15},
	publisher = {arXiv},
	author = {Nayana, A. J. and Margutti, Raffaella and Wiston, Eli and Laskar, Tanmoy and Migliori, Giulia and Chornock, Ryan and Galvin, Timothy J. and LeBaron, Natalie and Hajela, Aprajita and Christy, Collin T. and Sfaradi, Itai and Tsuna, Daichi and Aspegren, Olivia and De Colle, Fabio and Metzger, Brian D. and Lu, Wenbin and Beniamini, Paz and Kasen, Daniel and Berger, Edo and Grefenstette, Brian W. and Alexander, Kate D. and Anupama, G. C. and Coppejans, Deanne L. and Cruz, Luigi F. and DeBoer, David R and Drout, Maria R. and Farah, Wael and Huang, Xiaoshan and Jacobson-Galán, W. V. and Milisavljevic, Dan and Pollak, Alexander W. and Roth, Nathan J. and Sears, Huei and Siemion, Andrew and Sheikh, Sofia Z. and Steiner, James F. and Vurm, Indrek},
	month = aug,
	year = {2025},
}

@misc{cendes_continued_2025,
	title = {Continued {Rapid} {Radio} {Brightening} of the {Tidal} {Disruption} {Event} {AT2018hyz}},
	url = {https://ui.adsabs.harvard.edu/abs/2025arXiv250708998C},
	doi = {10.48550/arXiv.2507.08998},
	urldate = {2025-08-18},
	publisher = {arXiv},
	author = {Cendes, Yvette and Berger, Edo and Beniamini, Paz and Gill, Ramandeep and Matsumoto, Tatsuya and Alexander, Kate D. and Bietenholz, Michael F. and Hajela, Aprajita and Christy, Collin T. and Chornock, Ryan and Gomez, Sebastian and Gurwell, Mark A. and Keating, Garrett K. and Laskar, Tanmoy and Margutti, Raffaella and Rao, Ramprasad and Velez, Natalie and Wieringa, Mark H.},
	month = jul,
	year = {2025},
}

@article{beniamini_swift_2023,
	title = {Swift {J1644}+57 as an off-axis {Jet}},
	volume = {524},
	issn = {0035-8711},
	url = {https://ui.adsabs.harvard.edu/abs/2023MNRAS.524.1386B},
	doi = {10.1093/mnras/stad1950},
	urldate = {2025-05-04},
	journal = {Monthly Notices of the Royal Astronomical Society},
	publisher = {OUP},
	author = {Beniamini, Paz and Piran, Tsvi and Matsumoto, Tatsuya},
	month = sep,
	year = {2023},
	pages = {1386--1395},
}

@article{ryan_modeling_2024,
	title = {Modeling of {Long}-term {Afterglow} {Counterparts} to {Gravitational} {Wave} {Events}: {The} {Full} {View} of {GRB} {170817A}},
	volume = {975},
	issn = {0004-637X},
	shorttitle = {Modeling of {Long}-term {Afterglow} {Counterparts} to {Gravitational} {Wave} {Events}},
	url = {https://ui.adsabs.harvard.edu/abs/2024ApJ...975..131R},
	doi = {10.3847/1538-4357/ad6a14},
	urldate = {2025-08-31},
	journal = {The Astrophysical Journal},
	author = {Ryan, Geoffrey and van Eerten, Hendrik and Troja, Eleonora and Piro, Luigi and O'Connor, Brendan and Ricci, Roberto},
	month = nov,
	year = {2024},
	pages = {131},
}

@article{beniamini_afterglow_2020,
	title = {Afterglow light curves from misaligned structured jets},
	volume = {493},
	issn = {0035-8711},
	url = {https://ui.adsabs.harvard.edu/abs/2020MNRAS.493.3521B},
	doi = {10.1093/mnras/staa538},
	urldate = {2025-05-04},
	journal = {Monthly Notices of the Royal Astronomical Society},
	publisher = {OUP},
	author = {Beniamini, Paz and Granot, Jonathan and Gill, Ramandeep},
	month = apr,
	year = {2020},
	pages = {3521--3534},
}

@article{ryan_gamma-ray_2020,
	title = {Gamma-{Ray} {Burst} {Afterglows} in the {Multimessenger} {Era}: {Numerical} {Models} and {Closure} {Relations}},
	volume = {896},
	issn = {0004-637X},
	shorttitle = {Gamma-{Ray} {Burst} {Afterglows} in the {Multimessenger} {Era}},
	url = {https://ui.adsabs.harvard.edu/abs/2020ApJ...896..166R},
	doi = {10.3847/1538-4357/ab93cf},
	urldate = {2024-07-18},
	journal = {The Astrophysical Journal},
	publisher = {IOP},
	author = {Ryan, Geoffrey and van Eerten, Hendrik and Piro, Luigi and Troja, Eleonora},
	month = jun,
	year = {2020},
	pages = {166},
}

@article{ghirlanda_compact_2019,
	title = {Compact radio emission indicates a structured jet was produced by a binary neutron star merger},
	volume = {363},
	issn = {0036-8075},
	url = {https://ui.adsabs.harvard.edu/abs/2019Sci...363..968G},
	doi = {10.1126/science.aau8815},
	urldate = {2023-10-19},
	journal = {Science},
	author = {Ghirlanda, G. and Salafia, O. S. and Paragi, Z. and Giroletti, M. and Yang, J. and Marcote, B. and Blanchard, J. and Agudo, I. and An, T. and Bernardini, M. G. and Beswick, R. and Branchesi, M. and Campana, S. and Casadio, C. and Chassande-Mottin, E. and Colpi, M. and Covino, S. and D'Avanzo, P. and D'Elia, V. and Frey, S. and Gawronski, M. and Ghisellini, G. and Gurvits, L. I. and Jonker, P. G. and van Langevelde, H. J. and Melandri, A. and Moldon, J. and Nava, L. and Perego, A. and Perez-Torres, M. A. and Reynolds, C. and Salvaterra, R. and Tagliaferri, G. and Venturi, T. and Vergani, S. D. and Zhang, M.},
	month = mar,
	year = {2019},
	pages = {968--971},
}

@article{granot_shape_2002,
	title = {The {Shape} of {Spectral} {Breaks} in {Gamma}-{Ray} {Burst} {Afterglows}},
	volume = {568},
	issn = {0004-637X},
	url = {https://ui.adsabs.harvard.edu/abs/2002ApJ...568..820G},
	doi = {10.1086/338966},
	urldate = {2025-11-24},
	journal = {The Astrophysical Journal},
	publisher = {IOP},
	author = {Granot, Jonathan and Sari, Re'em},
	month = apr,
	year = {2002},
	pages = {820--829},
}

@article{waxman_gamma-ray--burst_1997,
	title = {Gamma-{Ray}--{Burst} {Afterglow}: {Supporting} the {Cosmological} {Fireball} {Model}, {Constraining} {Parameters}, and {Making} {Predictions}},
	volume = {485},
	issn = {0004-637X},
	shorttitle = {Gamma-{Ray}--{Burst} {Afterglow}},
	url = {https://ui.adsabs.harvard.edu/abs/1997ApJ...485L...5W},
	doi = {10.1086/310809},
	urldate = {2025-12-29},
	journal = {The Astrophysical Journal},
	publisher = {IOP},
	author = {Waxman, Eli},
	month = aug,
	year = {1997},
	pages = {L5--L8},
}

@article{meszaros_optical_1997,
	title = {Optical and {Long}-{Wavelength} {Afterglow} from {Gamma}-{Ray} {Bursts}},
	volume = {476},
	issn = {0004-637X},
	url = {https://ui.adsabs.harvard.edu/abs/1997ApJ...476..232M},
	doi = {10.1086/303625},
	urldate = {2025-11-24},
	journal = {The Astrophysical Journal},
	publisher = {IOP},
	author = {Mészáros, P. and Rees, M. J.},
	month = feb,
	year = {1997},
	pages = {232--237},
}

@article{wang_stability_2003,
	title = {Stability of an {Ultrarelativistic} {Blast} {Wave} in an {External} {Medium} with a {Steep} {Power}-{Law} {Density} {Profile}},
	volume = {594},
	issn = {0004-637X},
	url = {https://ui.adsabs.harvard.edu/abs/2003ApJ...594..924W},
	doi = {10.1086/377154},
	urldate = {2025-04-08},
	journal = {The Astrophysical Journal},
	publisher = {IOP},
	author = {Wang, Xiaohu and Loeb, Abraham and Waxman, Eli},
	month = sep,
	year = {2003},
	pages = {924--935},
}

@article{kushnir_closing_2010,
	title = {Closing the {Gap} in the {Solutions} of the {Strong} {Explosion} {Problem}: {An} {Expansion} of the {Family} of {Second}-type {Self}-similar {Solutions}},
	volume = {723},
	issn = {0004-637X},
	shorttitle = {Closing the {Gap} in the {Solutions} of the {Strong} {Explosion} {Problem}},
	url = {https://ui.adsabs.harvard.edu/abs/2010ApJ...723...10K},
	doi = {10.1088/0004-637X/723/1/10},
	urldate = {2025-03-10},
	journal = {The Astrophysical Journal},
	publisher = {IOP},
	author = {Kushnir, Doron and Waxman, Eli},
	month = nov,
	year = {2010},
	pages = {10--19},
}

@article{sedov_propagation_1946,
	title = {Propagation of strong shock waves},
	volume = {10},
	issn = {0021-8928},
	url = {https://ui.adsabs.harvard.edu/abs/1946JApMM..10..241S},
	urldate = {2023-08-17},
	journal = {Journal of Applied Mathematics and Mechanics},
	author = {Sedov, Leonid Ivanovich},
	month = jan,
	year = {1946},
	pages = {241--250},
}

@article{taylor_formation_1950,
	title = {The {Formation} of a {Blast} {Wave} by a {Very} {Intense} {Explosion}. {I}. {Theoretical} {Discussion}},
	volume = {201},
	issn = {0080-46301364-5021},
	url = {https://ui.adsabs.harvard.edu/abs/1950RSPSA.201..159T},
	doi = {10.1098/rspa.1950.0049},
	urldate = {2023-08-17},
	journal = {Proceedings of the Royal Society of London Series A},
	author = {Taylor, Geoffrey},
	month = mar,
	year = {1950},
	pages = {159--174},
}

@book{sedov_similarity_1959,
	address = {New York},
	title = {Similarity and {Dimensional} {Methods} in {Mechanics}},
	urldate = {2026-08-10},
	publisher = {Academic Press},
	author = {Sedov, L. I.},
	month = jan,
	year = {1959},
}

@incollection{von_neumann_point_1963,
	address = {Oxford},
	title = {The {Point} {Source} {Solution}},
	url = {https://ui.adsabs.harvard.edu/abs/1961cowo.book.....V},
	urldate = {2026-06-25},
	booktitle = {John von {Neumann}: {Collected} {Works}, {Vol}. 6: {Theory} of {Games}, {Astrophysics}, {Hydrodynamics} and {Meteorology}},
	publisher = {Pergamon Press},
	author = {von Neumann, J.},
	year = {1963},
	note = {John von Neumann: Collected Works, Vol. 6: Theory of Games, Astrophysics, Hydrodynamics and Meteorology},
	pages = {219--237},
}

@article{sadeh_hydrodynamics_2026,
	title = {The hydrodynamics of stratified ultra-relativistic outflows and the origin of {GRB} {X}-ray plateaus},
	issn = {0035-8711},
	url = {https://doi.org/10.1093/mnras/stag833},
	doi = {10.1093/mnras/stag833},
	urldate = {2026-05-07},
	journal = {Monthly Notices of the Royal Astronomical Society},
	author = {Sadeh, Gilad and Hotokezaka, Kenta and Shibata, Masaru},
	month = may,
	year = {2026},
	pages = {stag833},
}

@article{faran_non-relativistic_2021,
	title = {The non-relativistic interiors of ultra-relativistic explosions: {Extension} to the {Blandford}–{McKee} solutions},
	volume = {33},
	issn = {1070-6631},
	shorttitle = {The non-relativistic interiors of ultra-relativistic explosions},
	url = {https://doi.org/10.1063/5.0037299},
	doi = {10.1063/5.0037299},
	number = {2},
	urldate = {2026-03-27},
	journal = {Physics of Fluids},
	author = {Faran, Tamar and Sari, Re'em},
	month = feb,
	year = {2021},
	pages = {026105},
}

@article{sari_first_2006,
	title = {First and second type self-similar solutions of implosions and explosions containing ultrarelativistic shocks},
	volume = {18},
	issn = {1070-6631},
	url = {https://doi.org/10.1063/1.2174567},
	doi = {10.1063/1.2174567},
	number = {2},
	urldate = {2025-07-10},
	journal = {Physics of Fluids},
	author = {Sari, Re’em},
	month = feb,
	year = {2006},
	pages = {027106},
}

@article{pan_composite_2009,
	title = {Composite self-similar solutions for relativistic shocks: {The} transition to cold fluid temperatures},
	volume = {21},
	issn = {1070-6631},
	shorttitle = {Composite self-similar solutions for relativistic shocks},
	url = {https://doi.org/10.1063/1.3249751},
	doi = {10.1063/1.3249751},
	number = {11},
	urldate = {2025-07-06},
	journal = {Physics of Fluids},
	author = {Pan, Margaret and Sari, Re’em},
	month = nov,
	year = {2009},
	pages = {116101},
}

@article{best_second-type_2000,
	title = {Second-type self-similar solutions to the ultrarelativistic strong explosion problem},
	volume = {12},
	issn = {1070-6631},
	url = {https://doi.org/10.1063/1.1285921},
	doi = {10.1063/1.1285921},
	number = {11},
	urldate = {2025-03-25},
	journal = {Physics of Fluids},
	author = {Best, Paul and Sari, Re’em},
	month = nov,
	year = {2000},
	pages = {3029--3035},
}

@article{waxman_secondtype_1993,
	title = {Second‐type self‐similar solutions to the strong explosion problem},
	volume = {5},
	issn = {0899-8213},
	url = {https://doi.org/10.1063/1.858668},
	doi = {10.1063/1.858668},
	number = {4},
	urldate = {2025-03-25},
	journal = {Physics of Fluids A: Fluid Dynamics},
	author = {Waxman, Eli and Shvarts, Dov},
	month = apr,
	year = {1993},
	pages = {1035--1046},
}

@misc{gruzinov_self-similarity_2003,
	title = {Self-similarity of the {Third} {Type} in the {Strong} {Explosion} {Problem}},
	url = {http://arxiv.org/abs/astro-ph/0303242},
	doi = {10.48550/arXiv.astro-ph/0303242},
	urldate = {2025-03-17},
	publisher = {arXiv},
	author = {Gruzinov, Andrei},
	month = mar,
	year = {2003},
	note = {arXiv:astro-ph/0303242},
}

@article{blandford_fluid_1976,
	title = {Fluid dynamics of relativistic blast waves},
	volume = {19},
	issn = {0031-9171},
	url = {https://aip.scitation.org/doi/abs/10.1063/1.861619},
	doi = {10.1063/1.861619},
	number = {8},
	urldate = {2022-09-21},
	journal = {The Physics of Fluids},
	publisher = {American Institute of Physics},
	author = {Blandford, R. D. and McKee, C. F.},
	month = aug,
	year = {1976},
	pages = {1130--1138},
}

\appendix
\section{Glossary}
Table~\ref{tab:glossary} lists the main symbols used in the paper.
\label{app:glossary}
\begin{table*}[t]
\caption{\label{tab:glossary}Glossary of the main symbols.}
\begin{ruledtabular}
\begin{tabular}{ll}
Symbol & Definition \\
\hline
\multicolumn{2}{l}{\textit{Hot BM solution}}\\
\hline
$k$ & external density power-law index, $\rho_{\rm ext}\propto r^{-k}$ \\
\hline
$m$ & shock evolution exponent for the impulsive BM solution, $\Gamma^2\propto t^{-m}$, $m=3-k$ \\
\hline
$\Gamma$, $\gamma$ & Lorentz factors of the shock front and of the fluid \\
\hline
$\chi$ & BM similarity variable, $\chi\simeq\left[1+2(m+1)\Gamma^2\right](1-r/t)$ \\
\hline
$f,g,h$ & BM self-similar profiles of the pressure, Lorentz factor, and lab frame density \\
\hline
$\alpha_f,\alpha_h$ & BM profile exponents, $f=\chi^{\alpha_f}$, $h=\chi^{\alpha_h}$ \\
\hline
$\chi_{\rm slow}$ & kinematic trans-relativistic cutoff, $\chi_{\rm slow}=\Gamma^2/18$ \\
\hline
$\chi_{\rm cold}$ & thermodynamic trans-relativistic cutoff, defined by $p/\rho=1$ \\
\hline
$\chi_b$ & first-breakdown coordinate, $\min(\chi_{\rm slow},\chi_{\rm cold})$ \\
\hline
$\Gamma_{\rm launch},\Gamma_{\rm hit}$ & shock Lorentz factor when a signal is launched from $\chi_b$ and when it reaches the shock \\
\hline
$k_g$ & critical index for causal contact, $k_g=\tfrac{7\sqrt3}{2}-4\simeq2.062$ \\
\hline
\multicolumn{2}{l}{\textit{Cooling solution}}\\
\hline
$\delta$ & cooling-layer scale, $\delta=R-r_{\rm cold}$ \\
\hline
$\xi$ & cooling similarity variable, $\xi=(R-r)/\delta$ \\
\hline
$\overline\Gamma,\overline P,\overline N$ & cooling scales of the Lorentz factor, pressure, and density \\
\hline
$\overline f,\overline g,\overline h$ & cooling self-similar profiles of the pressure, Lorentz factor, and density \\
\hline
$q_\gamma,q_p,q_\delta$ & time exponents, $q_\gamma=d\ln\overline\Gamma/d\ln t$, $q_p=d\ln\overline P/d\ln t$, $q_\delta=d\ln\delta/d\ln t$ \\
\hline
$\lambda$ & drift velocity of a fluid element through the pattern, $\lambda=t\,d\xi/dt$ \\
\hline
$z$ & drift variable, $z=q_\delta\xi\overline g/[2(4-k)]$ \\
\hline
$\eta$ & temperature variable, $\eta=\overline f\sqrt{\overline g}/\overline h$ \\
\hline
$c_s$ & sound speed, $c_s^2=4\eta/[3(1+4\eta)]$ \\
\hline
$\mathcal D_+$ & forward-characteristic drift factor, $\mathcal D_+=(1-c_s)/(1+c_s)-z$ \\
\hline
$\overline f_\xi,\overline g_\xi,\overline h_\xi$ & logarithmic slopes, $\overline f_\xi=d\ln\overline f/d\ln\xi$, etc. \\
\hline
$\xi_s$ & location of the inner sonic point \\
\hline
$k_c$ & merger of the finite-$c_s$ roots, $k_c\simeq2.181$ \\
\hline
$\mu_\rho$ & endpoint density exponent, $\rho\propto(\xi_s-\xi)^{\mu_\rho}$, $\mu_\rho=(12-5k)/k$ \\
\hline
$\lambda_\eta,\lambda_z$ & overlap exponents of the EoS correction and of the free mode \\
\hline
$C$ & amplitude of the free mode \\
\end{tabular}
\end{ruledtabular}
\end{table*}

\section{$\frac{d\chi_\pm}{dt}$ derivation}
\label{app:a}
In terms of the BM self-similar coordinate, a forward-going characteristic is given by
\begin{equation}
\chi_\pm=\left[1+2(4-k)\Gamma^2\right]\left(1-\frac{r_\pm}{t}\right).
\end{equation}
Its derivative is
\begin{equation}
\begin{aligned}
     \frac{d\chi_\pm}{dt}&=\frac{1}{t}\left[1+2(4-k)\Gamma^2\right]\left(\frac{r_\pm}{t}-\frac{dr_\pm}{dt}\right)\\&-2\frac{(3-k)}{t}(4-k)\Gamma^2\left(1-\frac{r_\pm}{t}\right),\\
    &=\frac{1}{t}\left[1+2(4-k)\Gamma^2\right]\left(1-\frac{dr_\pm}{dt}\right)\\&-\frac{1}{t}\left(1-\frac{r_\pm}{t}\right)\left[1+2(4-k)\Gamma^2+2(3-k)(4-k)\Gamma^2\right],\\
    &=\frac{1}{t}\left[1+2(4-k)\Gamma^2\right]\left(1-\frac{dr_\pm}{dt}\right)\\&-\frac{\chi_\pm}{t}\left[4-k-\frac{3-k}{2(4-k)\Gamma^2}\right].
\end{aligned}
\end{equation}
Using Eq.~(\ref{eq:drdt}) we find
\begin{equation}
\begin{aligned}
\frac{d\chi_\pm}{dt}&=\frac{1}{t}\left[1+2(4-k)\Gamma^2\right]\left(\frac{\chi_\pm}{\Gamma^2}(\mp\sqrt{3}+2)\right)\\&-\frac{\chi_\pm}{t}\left[4-k-\frac{3-k}{2(4-k)\Gamma^2}\right],\\
    &=\frac{\chi_\pm}{t}\left[(3\mp2\sqrt{3})(4-k)+\frac{1}{\Gamma^2}\left(2\mp\sqrt{3}+\frac{3-k}{2(4-k)}\right)\right].
\end{aligned}
\end{equation}
Keeping the leading term in the ultra-relativistic limit, we obtain
\begin{equation}
\begin{aligned}
    \frac{d\chi_\pm}{dt}=\frac{\chi_\pm}{t}\left[(3\mp2\sqrt{3})(4-k)\right].
\end{aligned}
\end{equation}

\section{The causal criterion in planar and cylindrical geometry}
\label{app:geometry}
We label the geometry by $\alpha_g=0,1,2$ (planar, cylindrical,
spherical), replacing $r^{-2}\partial_r(r^2\,\cdot)$ by
$r^{-\alpha_g}\partial_r(r^{\alpha_g}\,\cdot)$ and $2p/r$ by $\alpha_g p/r$ in
Eqs.~\eqref{eq:relativistic_euler}. For an impulsive adiabatic blast
wave, energy conservation gives the first-type similarity
exponent\citep{sari_first_2006} $m=1+\alpha_g-k$, and with the ansatz of
Eq.~\eqref{eq:BM} the profiles remain power laws in $\chi$, with
$g=\chi^{-1}$ for all $\alpha_g$.\citep{sari_first_2006} The identity
$p/\rho=f\Gamma\sqrt g/(3\sqrt2\,h)$ is unchanged, and
\begin{equation}
\chi_{\rm cold}=\left(\frac{\Gamma^2}{18}\right)^{\frac{3(2+\alpha_g-k)}{k+\alpha_g+2}},
\end{equation}
while $\chi_{\rm slow}=\Gamma^2/18$ as before. The two cutoffs
exchange order at $k=1+\alpha_g/2$. The characteristic drift generalizes
Eq.~\eqref{eq:dchi_dt} with $4-k\to m+1$,
\begin{equation}
\frac{d\ln\chi_\pm}{d\ln t}=(3\mp2\sqrt3)\,(m+1).
\end{equation}
The causal boundary, at which
$d\ln\chi_+/d\ln t=d\ln\chi_b/d\ln t$, is
\begin{equation}
k=\begin{cases}
1+\alpha_g-\frac{\sqrt3}{2},&\text{(kinematic branch)},\\[2pt]
\left(\frac32+\alpha_g\right)\sqrt3-(2+\alpha_g),&\text{(thermodynamic branch)}.
\end{cases}
\end{equation}
On the kinematic branch, the condition is a statement about the
deceleration rate alone, $m=\sqrt3/2$, independent of geometry. The realized transition is the branch value lying inside its own domain
relative to $k=1+\alpha_g/2$: the kinematic one for $\alpha_g<\sqrt3$ and
the thermodynamic one for $\alpha_g>\sqrt3$. Hence
\begin{equation}
k_g(\alpha_g)=
\begin{cases}
1-\frac{\sqrt3}{2}\simeq0.134,& \alpha_g=0,\\
2-\frac{\sqrt3}{2}\simeq1.134,& \alpha_g=1,\\
\frac{7\sqrt3}{2}-4\simeq2.062,& \alpha_g=2 .
\end{cases}
\end{equation}
Only in the spherical case is the causal boundary set by the
thermodynamic transition, and the causal protection established in this paper is increasingly fragile in lower-dimensional geometry. For $\alpha_g<\sqrt3$ and $k>k_g$, the breakdown at $\chi_b$ is kinematic for $k_g<k<1+\alpha_g/2$ and thermodynamic for $k>1+\alpha_g/2$. In such a case, the cooling construction of \S~\ref{sec:cooling}, derived for spherical geometry, does not carry over, and the flow beyond $\chi_b$ requires the general-geometry counterparts of the cooling layer and of the trans-relativistic interior.\citep{faran_non-relativistic_2021} We defer this analysis, together with its implications for the planar first-type solutions, to future work.

\section{\(d\xi_\pm/dt\) derivation}
\label{app:b}
In terms of the cooling similarity coordinate, a characteristic is given by
\begin{equation}
\xi_\pm=\frac{R-r_\pm}{\delta}.
\end{equation}
Its derivative is
\begin{equation}
\frac{d\xi_\pm}{dt}
=
\frac{\dot R-\dot r_\pm}{\delta}
-\xi_\pm\frac{\dot\delta}{\delta}.
\end{equation}
Using the characteristic velocity expansion,
\begin{equation}
\dot r_\pm
=
1-
\frac{1}{\overline\Gamma^2\overline g}
\frac{1\mp c_s}{1\pm c_s}
+O(\overline\Gamma^{-4}),
\end{equation}
and \(\dot\delta/\delta=q_\delta/t\), we obtain
\begin{equation}
\frac{d\xi_\pm}{dt}
=
\frac{\dot R-1}{\delta}
+
\frac{1}{\delta\,\overline\Gamma^2\overline g}
\frac{1\mp c_s}{1\pm c_s}
-
\frac{q_\delta\xi_\pm}{t}
+O\left(\frac{1}{\delta\overline\Gamma^4}\right).
\end{equation}
The shock-velocity correction is subleading in the cooling-layer limit.
Indeed, since \(\dot R-1=O(\Gamma^{-2})\) and
\(t/\delta\simeq2(4-k)\overline\Gamma^2\), its contribution to
\(t\,d\xi_\pm/dt\) is
\begin{equation}
\frac{t}{\delta}(\dot R-1)
=
O\left(\frac{\overline\Gamma^2}{\Gamma^2}\right)
=
O(\chi_{\rm cold}^{-1}).
\end{equation}
Dropping this term and keeping the leading cooling-layer contribution
gives
\begin{equation}
t\frac{d\xi_\pm}{dt}
=
\frac{2(4-k)}{\overline g}
\frac{1\mp c_s}{1\pm c_s}
-q_\delta\xi_\pm .
\end{equation}
Using the definition of \(z\), this becomes
\begin{equation}
\frac{d\xi_\pm}{dt}
=
\frac{1}{t}
\frac{2(4-k)}{\overline g}
\left[
\frac{1\mp c_s}{1\pm c_s}
-z
\right].
\end{equation}

\section{The zero-pressure endpoint}
\label{app:corner}
\subsection{Pressure-gradient condition}
Taking $\overline{f}\to0$ in Eqs.~(\ref{eq:ode_cooling}) while keeping $\eta\overline{f}_\xi$
finite, they reduce to
\begin{subequations}
\begin{equation}
\overline{g}_\xi-(1-z)\overline{h}_\xi=\frac{(q_\gamma+q_p+2)z}{q_\delta},
\label{eq:zp1}
\end{equation}
\begin{equation}
2(2-z)\,\eta\overline{f}_\xi+(1-z)\overline{h}_\xi+\frac{z-3}{2}\overline{g}_\xi=-\frac{(q_p+2)z}{q_\delta},
\label{eq:zp2}
\end{equation}
\begin{equation}
2(1-2z)\,\eta\overline{f}_\xi+(1-z)\overline{h}_\xi-\frac{z+1}{2}\overline{g}_\xi=-\frac{(2q_\gamma+q_p+2)z}{q_\delta}.
\label{eq:zp3}
\end{equation}
\end{subequations}
These remain finite as $z\to1$ provided $(1-z)\overline{h}_\xi$ has a finite
limit. Evaluating them at $z=1$ gives
\begin{subequations}
\begin{equation}
\left[\overline{g}_\xi\right]_{\xi_s}-\left[(1-z)\overline{h}_\xi\right]_{\xi_s}=\frac{q_\gamma+q_p+2}{q_\delta},
\label{eq:zpe1}
\end{equation}
\begin{equation}
2\left[\eta\overline{f}_\xi\right]_{\xi_s}+\left[(1-z)\overline{h}_\xi\right]_{\xi_s}-\left[\overline{g}_\xi\right]_{\xi_s}=-\frac{q_p+2}{q_\delta},
\label{eq:zpe2}
\end{equation}
\begin{equation}
-2\left[\eta\overline{f}_\xi\right]_{\xi_s}+\left[(1-z)\overline{h}_\xi\right]_{\xi_s}-\left[\overline{g}_\xi\right]_{\xi_s}=-\frac{2q_\gamma+q_p+2}{q_\delta}.
\label{eq:zpe3}
\end{equation}
\end{subequations}
Subtracting Eq.~\eqref{eq:zpe3} from Eq.~\eqref{eq:zpe2} yields the pressure-gradient condition
\begin{equation}
\left[\frac{\xi\overline{f}'\sqrt{\overline{g}}}{\overline{h}}\right]_{\xi_s}
 =\frac{q_\gamma}{2q_\delta} .
\end{equation}

\subsection{Density profile near the endpoint}
The conditions at $z=1$ fix only $[\eta\overline{f}_\xi]_{\xi_s}$
(Eq.~\eqref{eq:corner_grad}) and the combination
$[\overline{g}_\xi]_{\xi_s}-[(1-z)\overline{h}_\xi]_{\xi_s}$
(Eq.~\eqref{eq:zpe1}); they do not determine $[\overline{g}_\xi]_{\xi_s}$ by
itself. The missing piece is the limiting ratio $\eta/(1-z)$ as
$\xi\to\xi_s$, fixed by substituting
$\eta\propto(1-z)$ into Eq.~\eqref{eq:autonomous} and matching the
divergent terms, such that $\eta/(1-z)\to(k-2)/(13-2k)$. Both sides of this matching condition are proportional to $k-3$, so
at $k=3$ it is satisfied by any ratio, and the corner admits a one-parameter family of solutions, with the corner values below independent of the ratio. Using Eqs.~(\ref{eq:uv})--(\ref{eq:autonomous}), this approach ratio leads to
\begin{subequations}
\begin{align}
\left[\overline{g}_\xi\right]_{\xi_s}&=\frac{-14k^2+71k-80}{4(4k^2-21k+20)},\\
\left[(1-z)\overline{h}_\xi\right]_{\xi_s}&=\frac{10k^2-89k+156}{4(4k^2-21k+20)} .
\label{eq:gxi_corner}
\end{align}
\end{subequations}
The density slope itself follows from the continuity relation, Eq.~\eqref{eq:zp1}, which holds also for finite pressure,
\begin{equation}
\overline{h}_\xi=\frac{\overline{g}_\xi-(q_\gamma+q_p+2)z/q_\delta}{1-z}.
\label{eq:hxi_corner}
\end{equation}
Near the endpoint $d(1-z)/d\ln\xi\to-(1+\overline{g}_\xi)$, so that
$1-z\to(1+[\overline{g}_\xi]_{\xi_s})(\xi_s-\xi)/\xi_s$. Hence
$\overline{h}_\xi\to[(1-z)\overline{h}_\xi]_{\xi_s}/(1-z)$ proportional to
$(\xi_s-\xi)^{-1}$. Integrating $\overline{h}_\xi=d\ln\overline{h}/d\ln\xi$
gives $\overline{h}\propto(\xi_s-\xi)^{\mu_\rho}$, and since $\overline{g}$ is
finite at the endpoint, $\rho\propto(\xi_s-\xi)^{\mu_\rho}$ with
\begin{equation}
\mu_\rho=-\frac{\left[(1-z)\overline{h}_\xi\right]_{\xi_s}}{1+\left[\overline{g}_\xi\right]_{\xi_s}}=\frac{12-5k}{k}.
\label{eq:murho_app}
\end{equation}
The exponent changes sign at $k=12/5$: the endpoint density vanishes for
$k_c<k<12/5$, is finite only at $k=12/5$, and diverges for $12/5<k<3$.

\end{document}